\documentclass[iop,apj]{emulateapj}
\slugcomment{Accepted for publication in the Astrophysical Journal, April 27$^{\rm th}$, 2014}

\usepackage{amsmath}
\def\13CO{\mbox{$^{13}$CO}}
\def\C18O{\mbox{C$^{18}$O}}
\newcommand{\kms}{\mbox{${\rm km~s^{-1}}$}}
\newcommand{\e}{\mbox{$^{-1}$}}
\newcommand{\ee}{\mbox{$^{-2}$}}

\def\simgt{\lower.5ex\hbox{$\; \buildrel > \over \sim \;$}}
\def\simlt{\lower.5ex\hbox{$\; \buildrel < \over \sim \;$}}

\begin{document}

\title{A Parametric Modeling Approach to Measuring the Gas Masses of Circumstellar Disks}
\author{Jonathan P. Williams and William M. J. Best}
\affil{Institute for Astronomy, University of Hawaii at Manoa, Honolulu, HI, 96822, USA; jpw,wbest@ifa.hawaii.edu}
\shorttitle{Gas Masses of Circumstellar Disks}
\shortauthors{Williams \& Best}

\begin{abstract}
The disks that surround young stars are mostly composed of molecular gas,
which is harder to detect and interpret than the accompanying dust.
Disk mass measurements have therefore relied on large and uncertain
extrapolations from the dust to the gas.
We have developed a grid of models to study the dependencies of isotopologue
CO line strengths on disk structure and temperature parameters and find that
a combination of \13CO\ and \C18O\ observations provides a robust measure
of the gas mass. We apply this technique to Submillimeter Array observations
of nine circumstellar disks and published measurements of six well studied
disks. We find evidence for selective photodissociation of \C18O\
and determine masses to within a factor of about three.
The inferred masses for the nine disks in our survey range from
$0.7-6\,M_{\rm Jup}$, and all are well below the extrapolation from
the interstellar medium gas-to-dust ratio of 100.
This is consistent with the low masses of planets found around such stars,
and may be due to accretion or photoevaporation of a dust-poor upper atmosphere.
However, the masses may be underestimated if there are more efficient
CO depletion pathways than those known in molecular clouds and cold cores.
\end{abstract}

\keywords{circumstellar matter — planetary systems: protoplanetary disks
— solar system: formation}

\section{Introduction}
All sun-like stars form accompanied by rotationally supported disks of gas
and dust, a natural consequence of the conservation of angular momentum during
a gravitationally driven contraction over more than two orders of magnitude
from molecular cloud cores to disk scales.  Exoplanet surveys have now
established that the most common end-state of these disks is planets,
typically with sizes and masses substantially less than that of Jupiter
\citep{2013Sci...340..572H}.
A key first step toward understanding the origin and diversity of
exoplanetary systems, and thereby the formation of the Solar System,
is measuring circumstellar disk properties.  The most fundamental quantity
is mass, as this sets the amount of available raw material for forming planets.

Disks form with an initial composition inherited from the interstellar medium
(ISM) where the gas-to-dust mass ratio is 100 \citep{1978ApJ...224..132B}.
Despite the dust being such a minor constituent, it is the most
readily observed as it radiates over a continuum from infrared to millimeter
wavelengths.
The standard practice has been to convert this emission to dust mass
and then to total disk masses by assuming an ISM gas-to-dust ratio,
a large and uncertain extrapolation of two orders of magnitude.
Whereas the validity of this assumption has been confirmed on very large
scales in the case of molecular clouds, where alternative and independent
mass estimates can be made \citep{2001ApJ...547..792D},
this is a major source of uncertainty for circumstellar disks.
In particular, it is known that the dust grains grow from sub-micron
to millimeter sizes and beyond \citep{2005A&A...434..971D},
and such large grains are both structurally and thermally decoupled
from the gas \citep{2004ApJ...615..991K, 2006ApJ...638..314D}.

Most of the gas in a giant planet-forming disk is expected to be molecular
and relatively cool, $T < 100$\,K.  Collisional energies are too low for
the bulk constituent, H$_2$, to emit significantly so asymmetric molecules
with large dipole moments best trace the gas.  HD is the closest chemical
counterpart and was observed with far-infrared spectroscopy by the
Herschel Space Observatory in the closest known, gas-rich,
disk around TW\,Hydra \citep{2013Natur.493..644B}.
As Herschel is no longer operating, additional disk mass measurements using
this technique are not possible.  CO is the most abundant molecule
after H$_2$ and its millimeter wavelength rotational lines are the strongest
that are observable from the ground
\citep{2004A&A...425..955T, 2005MNRAS.359..663D}.
The CO lines are optically thick, however, and therefore an unreliable
measure of mass.
Even \13CO\ lines may have significant optical depth
\citep{1996A&A...309..493D, 2001A&A...377..566V}
but are less saturated and, together with constraints on \C18O,
provide a means to more accurately measure the
total number of molecules, as has been previously carried out for
molecular clouds and cores \citep{1997ApJ...491..615G}.

In this paper, we use Submillimeter Array (SMA) observations
of the $J=2-1$ transitions of CO, \13CO, and \C18O\ to constrain
the gas masses of nine circumstellar disks in the nearby Taurus cloud.
To analyze these data,
we develop a parametric model of disk gas density, temperature, and
chemistry to produce a large grid of CO isotopologue line luminosities.
We find that the luminosity of low lying rotational transitions of
\13CO\ and \C18O\ correlate with gas mass and that
we can determine its value to within at least an order of magnitude
from combinations of the two.
Although this is not as direct as the oft-used  formulaic approach to measuring
dust masses, this model grid approach provides a simple but robust way
to measure of the gas content of circumstellar disks that is suitable for
interpreting CO isotopologue line surveys in the future.

The observations are presented in \S\ref{sec:observations}.
The modeling is described in \S\ref{sec:modeling}, and we discuss
general insights regarding the CO emitting region in disks and
the utility of CO isotopologue lines for measuring disk masses.
We apply the parametric model grid to published observations
of well studied disks and the SMA survey in \S\ref{sec:results}.
Assuming that the CO-to-H$_2$ abundance is the same in disks as
in molecular clouds and cores, we find that the surveyed disks all have
low masses and lower gas-to-dust ratios than the ISM.
We discuss possible explanations and implications in
\S\ref{sec:discussion}, and conclude in \S\ref{sec:summary}.

\section{Observations}
\label{sec:observations}
We observed nine Class II disks in the Taurus star forming region.
This is one of the closest regions containing many young stars and has
been extensively studied \citep{2007prpl.conf..329G}.
The sources are all optically visible, single stars of spectral type K or M,
with large photospheric
excesses at all infrared wavelengths, indicative of full dusty disks
extending from the dust sublimation radius outwards.
Based on their luminosity and effective temperature,
the inferred stellar masses and ages are
$\sim 0.6-0.7\,M_\odot$ and $\sim 1-3$\,Myr, respectively.

The observations were carried out with the SMA on Maunakea, Hawaii.
The array consists of eight 6-m antenna configured
variously in compact and extended configurations to provide projected
baselines ranging from 5\,m to 240\,m.
Observations were carried out over 19 nights from November 2010 to
January 2013, in moderately dry weather conditions with precipitable
water vapor levels ranging from 1\,mm to 4\,mm,
corresponding to zenith atmospherical optical depths of
$0.05-0.15$ at the observing frequency of 230\,GHz (1.3\,mm wavelength).
A table of observational details is provided in Appendix~\ref{sec:obslog}.

The correlator was configured to place the $J=2-1$ lines of \C18O\
%(219.560358\,GHz)
and \13CO\ 
%(220.398684\,GHz)
in the lower sideband, and CO
%(230.538000\,GHz)
in the upper sideband.
High frequency resolution windows, consisting of 512 channels
of width 203\,kHz, were placed around these line
frequencies to provide a velocity resolution of 0.28\,\kms.
The cumulative continuum bandwidth away from the lines was 3.5\,GHz,
which provided high dust mass sensitivity.

The observations for any given night cycled through between 2 and 4 sources
and two gain calibrators (3C111, J0510+180) each half hour.
This provided sufficient calibration of the atmospheric amplitude and
phase variation and a wide range of spatial frequencies for each source
to provide high image fidelity.  The absolute flux scale was determined
by observations of an unresolved or at most marginally resolved planet
or giant planet satellite, variously Uranus, Callisto, or Titan.
Based on the variation of the amplitude calibration and flux calibration
measurements, we estimate the uncertainty on the source flux densities
to be 20\%.

\begin{figure*}[!ht]
\centering
\includegraphics[width=3.2in]{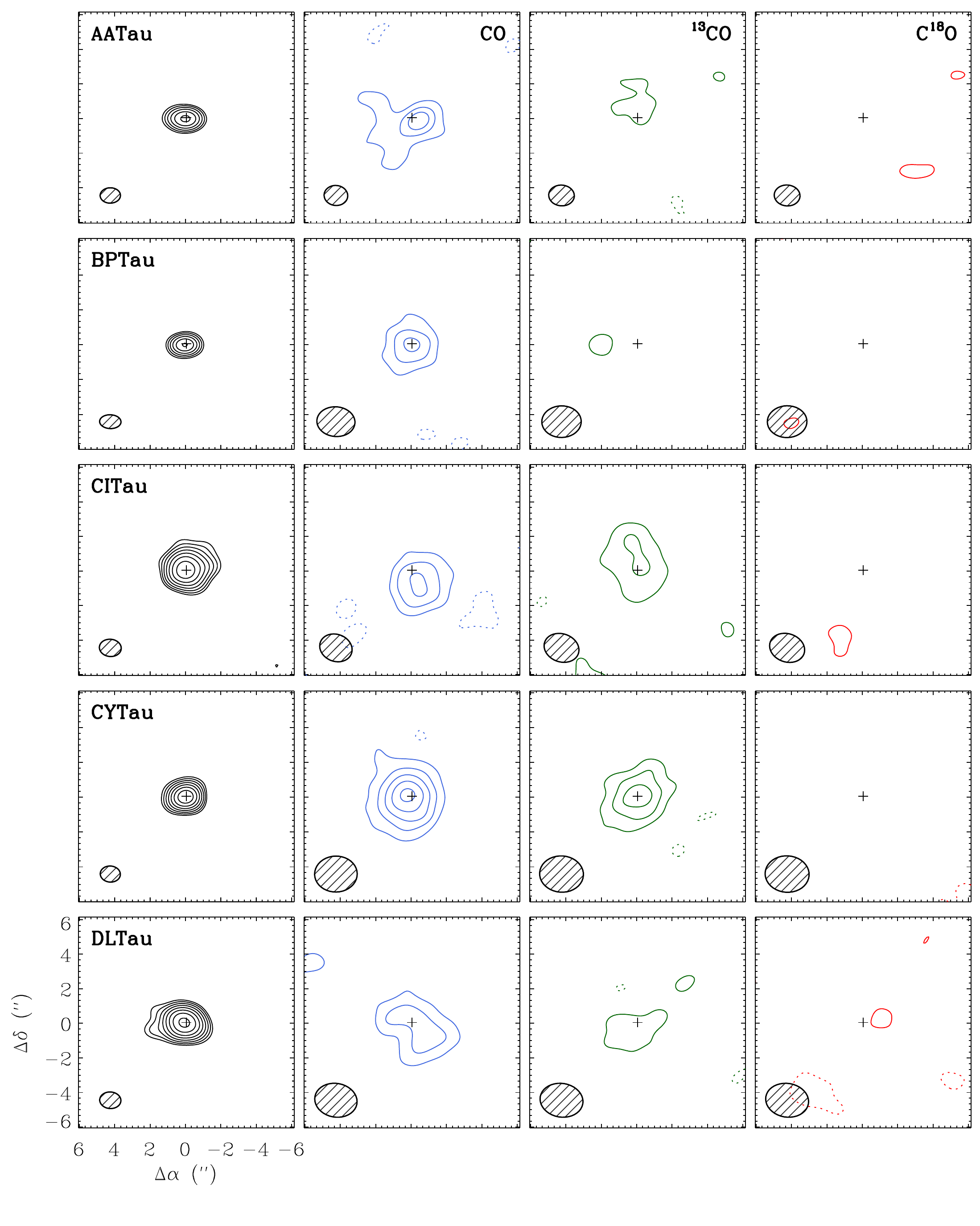}
\includegraphics[width=3.2in]{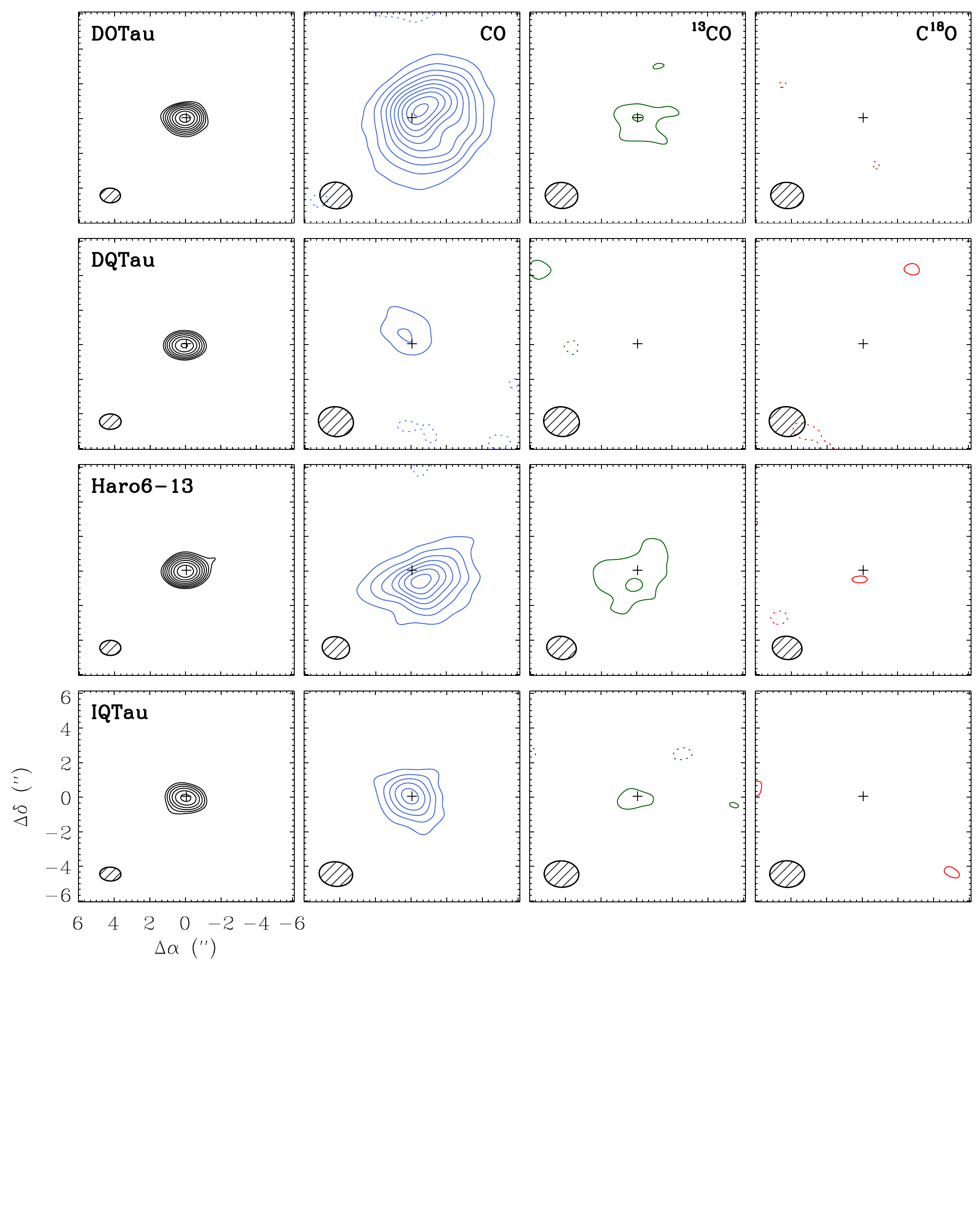}
\caption{
SMA maps of the continuum and integrated line intensities for the nine
Taurus disks in the survey. The black contours on the left hand side of each
panel show the continuum emission with contours beginning at 5\,mJy\,beam\e\
and increasing in multiples of 1.5. The integrated line maps have colored
contours, blue for CO, green for \13CO, and red for \C18O.
The contours are linear beginning at and increasing in steps of three
times the rms noise level, $\sigma$, listed in Table~\ref{tab:fluxes}.
The hashed ellipse shows the beamsize for each map in the lower left corner.
}
\label{fig:sma}
\end{figure*}

Images were produced from the calibrated visibilities using standard
inversion and cleaning procedures.  The typical resolution (beamsize)
of the final maps was $1\farcs 2\times 0\farcs 9$ for the uniformly
weighted continuum maps and $\sim 50$\% larger for the naturally
weighted line maps, the choice of weighting based on the
signal-to-noise in the data.
Figure~\ref{fig:sma} presents the continuum and velocity integrated
line maps for each source.

Whereas the continuum and CO emission are strongly detected in all
sources, and generally resolved, the much weaker \13CO\ and \C18O\ lines
were not detected with sufficient dynamic range to study their
spatial structure.  However, we found that we can determine gas
masses accurately from their integrated emission alone.
The integrated line intensities and rms noise levels for the
continuum and velocity integrated line maps are tabulated
in Table~\ref{tab:fluxes}.
In the case of a non-detection, we use the detected line, \13CO\
where possible else CO, to define the spatial and velocity range
over which to calculate the ($3\sigma$) upper limit to the emission.

\begin{deluxetable*}{lcccccccc}
\tablecolumns{9}
\tabletypesize{\tiny}
\tablewidth{0pt}
\tablecaption{SMA 1.3\,mm fluxes\label{tab:fluxes}}
\tablehead{
\colhead{} &
\multicolumn{2}{c}{$F_{\rm cont}$ (mJy)} &
\multicolumn{2}{c}{$F_{\rm CO}$ (Jy\,\kms)} &
\multicolumn{2}{c}{$F_{\rm ^{13}CO}$ (Jy\,\kms)} &
\multicolumn{2}{c}{$F_{\rm C^{18}O}$ (Jy\,\kms)} \\
\colhead{Source} &
\colhead{Value} & \colhead{$\sigma$} &
\colhead{Value} & \colhead{$\sigma$} &
\colhead{Value} & \colhead{$\sigma$} &
\colhead{Value} & \colhead{$\sigma$}
}
\startdata
AA Tau	        &  54.8	& 0.7 &  5.33 & 0.11 &   1.26  & 0.09 & $<0.27$ & 0.09 \\
BP Tau	        &  42.5	& 1.0 &  1.11 & 0.08 & $<0.24$ & 0.08 & $<0.21$ & 0.07 \\
CI Tau	        & 152	& 1.1 &  2.51 & 0.13 &   2.70  & 0.13 & $<0.39$ & 0.13 \\
CY Tau	        & 102	& 1.0 &  2.02 & 0.08 &   0.81  & 0.06 & $<0.15$ & 0.05 \\
DL Tau	        & 164	& 1.1 &  1.98 & 0.12 &   0.43  & 0.06 & $<0.15$ & 0.05 \\
DO Tau	        & 113	& 0.8 & 63.7  & 0.38 &   1.52  & 0.10 & $<0.30$ & 0.10 \\
DQ Tau	        &  62.6	& 0.7 &  0.97 & 0.11 & $<0.24$ & 0.08 & $<0.21$ & 0.07 \\
Haro 6-13       & 140	& 0.9 & 17.6  & 0.17 &   3.95  & 0.14 &   0.47  & 0.10 \\
IQ Tau	        &  57.6	& 1.0 &  2.52 & 0.08 &   0.48  & 0.07 & $<0.21$ & 0.07 \\[-2mm]
\enddata
%\tablenotetext{Non-detections are listed as $3\sigma$ upper limits.}
\end{deluxetable*}

\section{Modeling}
\label{sec:modeling}
Circumstellar disks are relatively small, faint objects that have, to date,
required long integrations with millimeter wavelength interferometers
to study their molecular gas content.
Consequently, only a small number of
disks have been imaged in isotopologue lines and most analyses
have been tailored to the individual object.
Driven by the moderately large sample size but low signal-to-noise
level in our data here, we use a different approach.  Rather than
analyze each disk individually, we create a large grid of models
that span a wide range of disk parameters, particularly in gas mass,
and compare with the data in a uniform way.  This approach is similar
to the SED modeling of young stellar objects by \citet{2006ApJS..167..256R}
and is also motivated by the grid modeling of mostly fine structure,
far-infrared lines of atomic species for comparison with Herschel
observations \citep{2010MNRAS.405L..26W, 2011A&A...532A..85K}.

\subsection{A parametric disk model}
\label{sec:parametric_disk_model}
\subsubsection{Density structure}
\label{sec:density_structure}
The basic model for the gas structure is an exponentially
tapered accretion disk profile in hydrostatic equilibrium
\citep{2011ARA&A..49...67W}.
This has its basis in the works of
\citet{2008ApJ...678.1119H} and \citet{2009ApJ...700.1502A},
who followed theoretical descriptions by
\citet{1974MNRAS.168..603L} and \citet{1998ApJ...495..385H}.

We define the azimuthally symmetric gas density $\rho(r,z)$
and temperature $T(r,z)$ in cylindrical coordinates.
For a given temperature structure (see below), we can determine
the shape of the vertical density structure by integrating
the equation of hydrostatic equilibrium,
\begin{equation}
\frac{\partial\ln\rho}{\partial z} =
  -\left[\left(\frac{GM_{\rm star}z}{(r^2+z^2)^{3/2}}\right)
   \left(\frac{\mu m_{\rm H}}{kT}\right)
  +\frac{\partial\ln T}{\partial z}\right],
\end{equation}
where $\mu = 2.37$ is the mean molecular weight of the gas and $m_{\rm H}$
is the mass of a hydrogen atom.  The resulting vertical profile is normalized
at each radius to have a vertically integrated surface density appropriate for
an accretion disk around a central gravitating point source,
\begin{equation}
\Sigma(r)=\Sigma_0\left(\frac{r}{r_c}\right)^{-\gamma}\,
          \exp\left[-\left(\frac{r}{r_c}\right)^{2-\gamma}\right].
\end{equation}

The global normalization is to the gas mass via
\begin{equation}
\Sigma_0=(2-\gamma)\,\frac{M_{\rm gas}}{2\pi r_c^2}
          \exp\left(\frac{r_{\rm in}}{r_c}\right)^{2-\gamma},
\end{equation}
where $r_{\rm in}$ is the inner radius of the disk.
We set this to 1\,AU as the vast majority of the gas mass
and of the CO millimeter emission is at large radii
and this allows the computational resources to be
concentrated on the outer disk. Our preliminary tests showed
that the models are not sensitive to this value (as long as it is small).

\begin{figure}[tb]
\centering
\includegraphics[width=3.4in]{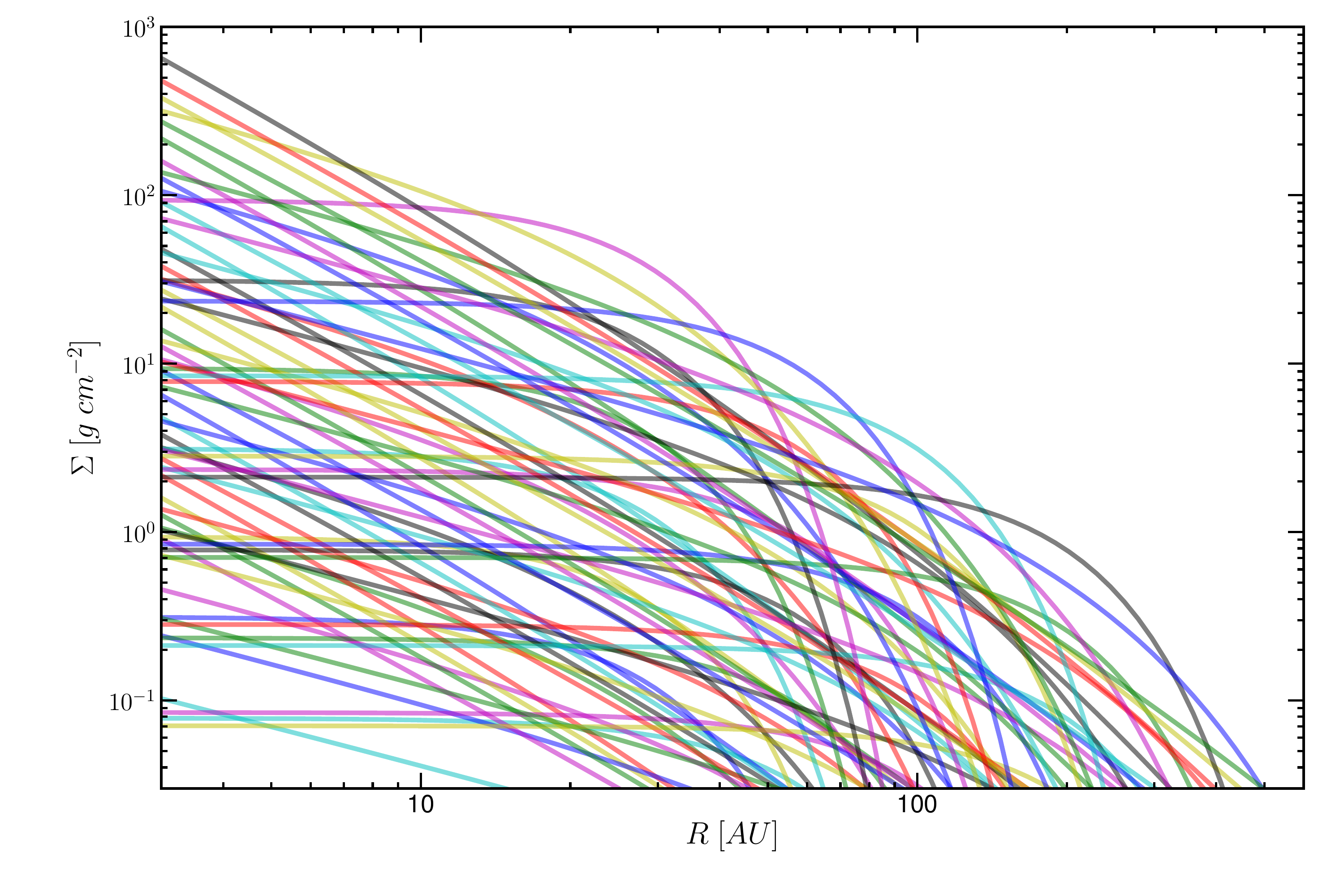}
\caption{
The range of surface density profiles in the model grid
for the range of parameters, \{$M_{\rm gas}, r_c, \gamma$\},
specified in Table~\ref{tab:grid}.
}
\label{fig:Sigma}
\end{figure}

For a given stellar mass,
this prescription for the disk physical structure has three
parameters; \{$M_{\rm gas}, r_c, \gamma$\}.
Figure~\ref{fig:Sigma} plots all the surface density profiles in
our model grid.
These encompass the range inferred (and extrapolated from dust
observations) for the most massive protoplanetary disks in
Ophiuchus observed by \citet{2009ApJ...700.1502A}
but extends to much lower masses, smaller sizes, and
also includes flatter profiles.

To calculate the radiative transfer, we also require the velocity
field, which we assume to be Keplerian with a Doppler width equal to
the sum of a thermal and turbulent component,
\begin{equation}
v^2(r,z) = \frac{GM_{\rm star}}{(r^2+z^2)^{1/2}}, \hspace{0.25in}
\Delta v_{\rm D}^2(r,z) = \frac{2kT}{\mu m_{\rm H}} + \Delta v_{\rm turb}^2.
\end{equation}
The prescription for the temperature is given below.
The turbulent component is known to be subsonic \citep{2011ApJ...727...85H},
and we fix its value at $\Delta v_{\rm turb}=0.01$\,\kms.

The vertical profile and rotational speed also depend on the
stellar mass and we include two values here,
$M_{\rm star} = 0.5, 1\,M_\odot$ that bracket the range in our
sample, to assess its effect on the CO line emission.

\subsubsection{Temperature structure}
\label{sec:temperature_structure}
The disk temperature structure is purely parametric, but grounded in
theory and observations.
%Disk accretion can be an important source of heating at small
%radii but stellar radiation dominates beyond about 2\,AU (D'Alessio 1998).
Deep in the disk, at $A_{\rm V}\simgt 10$\,mag, the densities are
sufficiently high for the gas and dust to be thermally coupled.
Fits to continuum observations show that the
radial midplane temperature gradient is a power law,
\begin{equation}
T_{\rm mid}(r) = T_{\rm mid,1}\left(\frac{r}{1\,{\rm AU}}\right)^{-q},
\end{equation}
with typical values, $T_{\rm mid,1} \simeq 200$\, K,
$q\simeq 0.55$, for Taurus protostars similar to those
in our CO survey \citep{2009ApJ...700.1502A}.

The temperature increases with height above the midplane due to
heating from scattered stellar UV photons at $A_{\rm V}\simeq 1-10$
\citep{2002A&A...389..464D, 2006ApJ...638..314D, 2011ApJ...740..118L}.
Furthermore, as the density decreases with increasing scale height, the gas and dust
thermally decouple.  Detailed models of the heating and cooling processes
show the gas temperature smoothly transitions from the midplane value
to a hotter atmospheric value at $z/r \sim 0.1-0.5$ for $r = 10-100$\,AU
\citep{2004ApJ...615..991K, 2007ApJ...661..334N, 2008ApJ...683..287G, 2009A&A...501..383W}.
The uppermost disk surface layer, at $A_{\rm V}\simlt 0.1$\,mag, may be
superheated to still higher temperatures but CO molecules would not
survive dissociation at such low column densities and we do not
include this feature in our models.

Following \citet{2003A&A...399..773D},
we parameterize the atmospheric temperature
as a radial power law in the same way as for the midplane profile,
\begin{equation}
T_{\rm atm}(r) = T_{\rm atm,1}\left(\frac{r}{1\,{\rm AU}}\right)^{-q}.
\end{equation}
As with \citet{2013ApJ...774...16R}, however,
we use a sine instead of cosine function in the connecting function between
the midplane and atmosphere so as to better match the vertical gas
temperature profiles with the above mentioned models,
\begin{equation}
T(r,z) =
  \begin{cases}
  T_{\rm mid} + (T_{\rm atm} - T_{\rm mid})
              \left[\sin\left(\frac{\pi z}{2z_q}\right)\right]^{2\delta}
              & \mbox{if }z < z_q\\
  T_{\rm atm} & \mbox{if } z\geq z_q
  \end{cases}.
\label{eq:temperature}
\end{equation}
This introduces two new parameters, $\delta, z_q$,
that describe the steepness of the profile and the height at
which the disk reaches the atmospheric value.
We fix $\delta=2$ as this provides a good approximation to
the gradient of the aforementioned theoretical models,
and $z_q=4H_{\rm p}$
where $H_p$ is the pressure scale height derived from the midplane
temperature, $(kT_{\rm mid}r^3/GM_{\rm star}\mu m_{\rm H})^{1/2}$,
based on the level at which the optical depth to
UV photons becomes unity \citep{2011ApJ...740..118L}.
These same values were also used by
\citet{2003A&A...399..773D} and \citet{2013ApJ...774...16R}.
We run the full grid for these fixed values but investigate their
effect on the line luminosities in Appendix~\ref{sec:temperature}.
There we show that they do not significantly affect our conclusions
or mass estimates drawn from the full grid.

\begin{figure}[tb]
\centering
\includegraphics[width=3.4in]{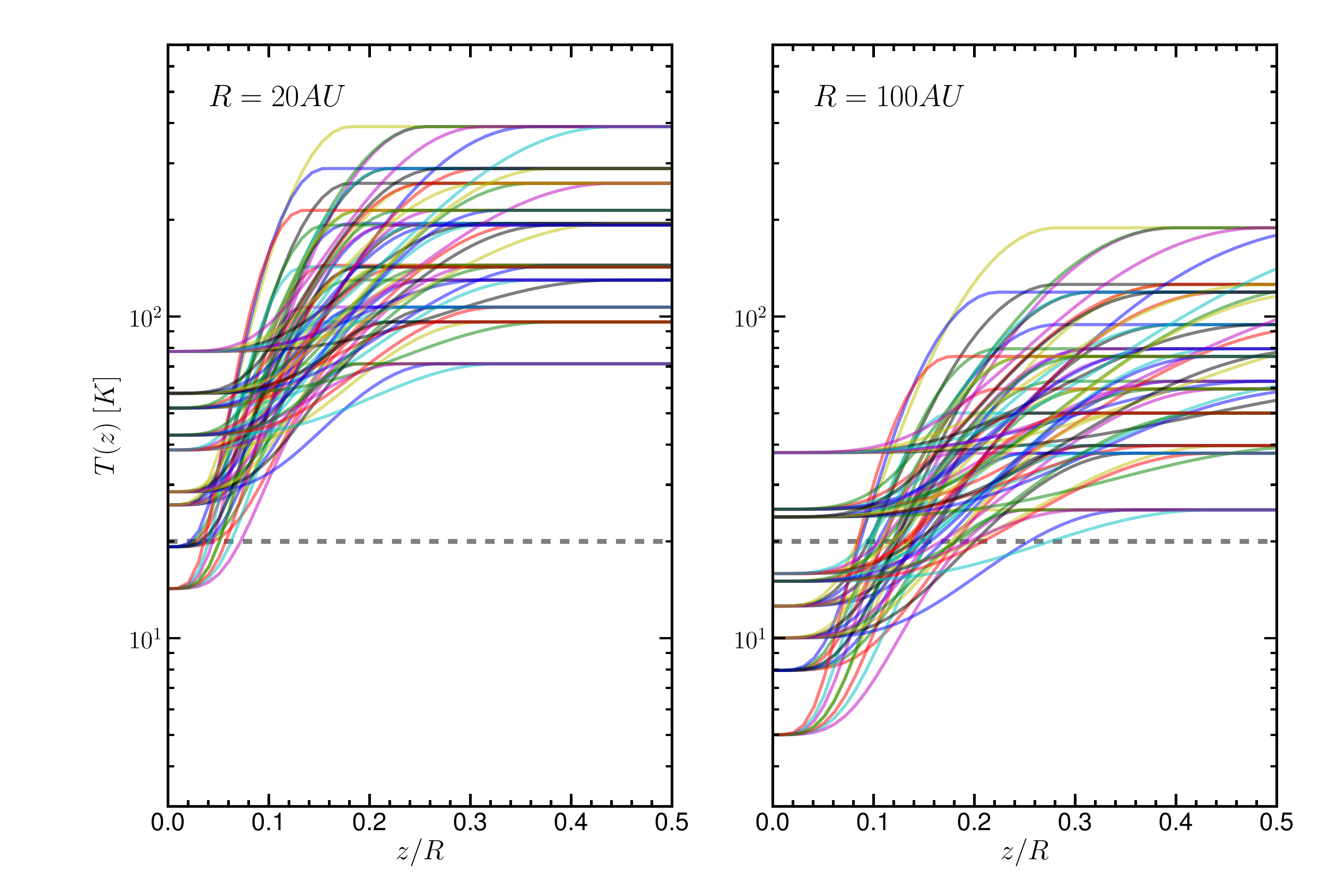}
\caption{
The range of vertical temperature density profiles at radii of 20\,AU
(left panel) and 100\,AU (right panel) in the model grid
for the range of parameters,
\{$M_{\rm star}, T_{\rm mid,1}, T_{\rm atm,1}, q$\},
specified in Table~\ref{tab:grid}.
The horizontal dashed line is the CO freeze-out temperature, 20\,K.
}
\label{fig:Tgas}
\end{figure}

This prescription for the disk temperature structure has three
parameters; \{$T_{\rm mid,1}, T_{\rm atm,1}, q$\}
and a dependence on stellar mass through $z_q$.
Figure~\ref{fig:Tgas} plots all the temperature profiles in our model
grid for two radii. These profiles bracket the range and have a similar
form as the aforementioned theoretical results.
Although our focus here is on T-Tauri stars, these profiles are
also sufficiently broad to encompass much of the expected range for
Herbig Ae stars \citep{2007A&A...463..203J}.

\subsubsection{CO chemistry}
\label{sec:CO_chemistry}
CO has a relatively simple chemistry that is readily incorporated into
our parametric model.  CO forms quickly in the gas phase and uses up all
the available carbon \citep{1988ApJ...334..771V}.
It is a very stable molecule with just two main destruction mechanisms,
freeze-out onto dust grains at low temperatures near the disk midplane
and photo-dissociation in the upper disk atmosphere.  These have each been
well characterized through several detailed studies.

CO depletion via freeze-out onto dust grains was first characterized
in molecular cores \citep{1999ApJ...523L.165C, 2002ApJ...569..815T, 2002A&A...389..908J},
then inferred in circumstellar disks
\citep{2008ApJ...681.1396Q, 2011ApJ...740...84Q},
and now directly resolved \citep{2013ApJ...774...16R}.
The resulting CO ``snowline'' also manifests itself in the enhanced
abundance of N$_2$H$^+$ and DCO$^+$
\citep{2013Sci...341..630Q, 2013A&A...557A.132M}.
From these results, we parameterize the freeze-out region as being
at temperatures, $T < 20$\,K.

Energetic radiation from the central star and the interstellar
radiation field dissociate CO in the upper layers of the disk.
This results in a sharp transition from C to CO that we set
to a column density,
$N_{\rm dissoc} = 1.3\times 10^{21}\,{\rm H}_2$\,cm\ee\
based on theory by \citet{2009A&A...503..323V}
and observations by \citet{2011ApJ...740...84Q}.

\begin{figure}[tb]
\centering
\includegraphics[width=3.5in]{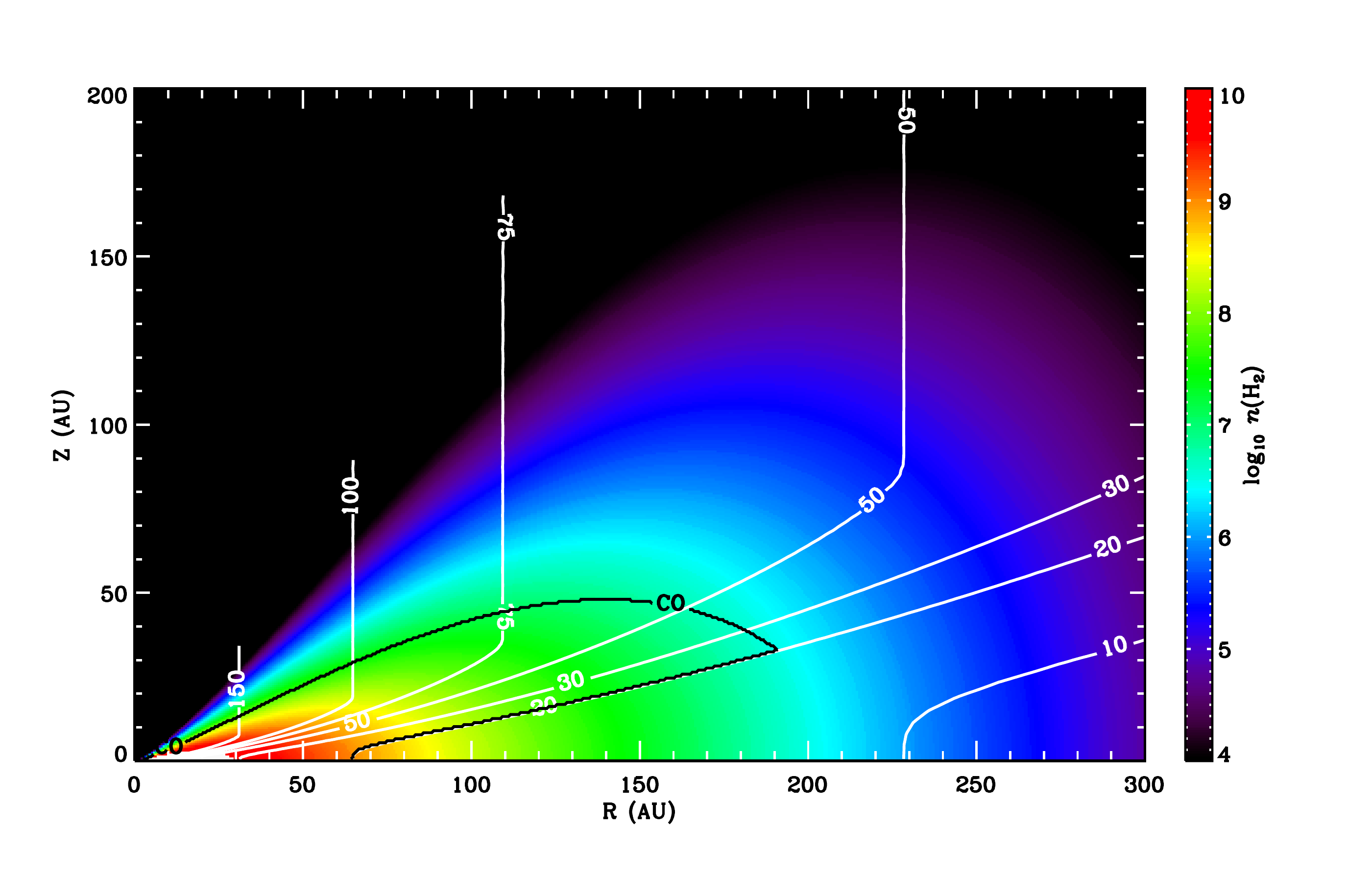}
\caption{
The density and temperature distribution of a model disk. The star is
at the origin and the disk is
radially symmetric with mirror symmetry about the midplane at $Z = 0$.
The color scale represents the H$_2$ gas density on a logarithmic scale.
The gas temperature is shown and labeled in the white contours.
The black contours, labeled ‘CO’, represent the boundary of a warm
molecular layer within which CO is expected to be in the gas phase
and emit millimeter wavelength rotational lines.
The parameters used for this model are $M_{\rm star}=1\,M_\odot,
M_{\rm gas}=0.01\,M_\odot, R_{\rm c}=60\,{\rm AU}, \gamma=0.75,
T_{\rm mid,1}=200\,{\rm K}, T_{\rm atm,1}=1000\,{\rm K}, q=0.55$.
}
\label{fig:disk_model}
\end{figure}

Consequently, a warm molecular layer, with nearly constant CO abundance,
lies between the depleted midplane and dissociation surface
\citep{2002A&A...386..622A, 2009A&A...501..383W}.
Our models incorporate these results as follows,
\begin{equation}
x({\rm CO}) = {\rm [CO]/[H_2]} =
  \begin{cases}
  1\times 10^{-4} &
    \mbox{$T > 20\,{\rm K}, N_{\rm H_2} > N_{\rm dissoc}$}\\
  0    & \mbox{elsewhere.}
  \end{cases}
\end{equation}
%This is the canonical value for dark clouds \citep{1978ApJS...37..407D},
%though actually derived from observations of \13CO.
An example disk density and temperature structure
is shown in Figure~\ref{fig:disk_model}.
The CO emitting region that we observe with our millimeter wavelength
observations is outlined by the black contours.

The isotopologues, \13CO\ and \C18O, have the same freeze-out temperature
as CO and will be depleted in the midplane. We therefore first model these
species with the same density profile as CO but scaled by their respective
isotopologue ratios,
[CO]/[\13CO]=70 and [CO]/[\C18O]=550 \citep{1994ARA&A..32..191W}.
These imply \13CO\ and \C18O\ abundances relative to H$_2$ 
that are consistent with cloud measurements by
\cite{2013MNRAS.431.1296R} and \cite{1982ApJ...262..590F} respectively.

The effect of photodissociation on the isotopologues is different than CO,
however, as higher total column densities are required for these rarer species
to self-shield.
\cite{2009A&A...503..323V} examined the complexities of this process,
including the effects of dust, H, H$_2$, CO mutual shielding, excitation,
doppler broadening, and ion-molecule isotope exchange reactions.
They find that the [CO]/[\13CO] ratio is relatively constant at the ISM
value in disks with moderate grain growth but that \C18O\
(and the rarer species, C$^{17}$O and $^{13}$C$^{18}$O)
should decrease steadily to a factor of $\sim 7$ lower abundance
relative to CO at high column densities toward the freeze-out region.
The particular density profile of \C18O\ within the
warm molecular layer at intermediate column densities,
$A_{\rm V}\simeq 1-10$\,mag, is therefore a computationally
complex function of both radius and height.
To minimize the number of additional parameters, and in keeping
with the simplicity of our approach, we approximate the effect
of selective photodissociation by calculating a second set of \C18O\ line
intensities with the isotopologue abundance reduced by a factor of 3,
[CO]/[\C18O]=1650.
The resulting two values, [\13CO]/[\C18O] $=8, 24$,
bracket recent observational results for molecular clouds in Orion
\citep{2014A&A...564A..68S}.

We emphasize that the model assumptions for the gas structure are only
azimuthal symmetry and hydrostatic equilibrium.
This allows us to consider fits to spectral line data independently
from fits to continuum data and therefore to infer gas properties
independently from those of the dust.
This approach is necessary to match the reality of these two
components being structurally, thermally, and dynamically decoupled.
In particular,
our constraints on gas masses are independent of dust mass
measurements or, within the confines of our modeling parameters,
any features in the dust structure such as
inner holes or strong azimuthal features \citep{2013Sci...340.1199V}.

\subsubsection{Radiative transfer}
\label{sec:radiative_transfer}
We calculate integrated line intensities for the models using 
the radiative transfer code RADMC-3D\footnotemark
\footnotetext{http://www.ita.uni-heidelberg.de/$\sim$dullemond/software/radmc-3d/}.
The gas density, temperature, and CO abundance profiles were
interpolated onto a spherical coordinate grid
with 100 logarithmically spaced points in radial distance from
1\,AU to 600\,AU and 60 linear steps in polar angle,
$\theta = 10^\circ$ to $90^\circ$, with axial and mirror symmetry.
We assumed Local Thermodynamic Equilibrium, which is a good
approximation for the low lying CO rotational states at the
densities and temperatures of the warm molecular region
\citep{2007ApJ...669.1262P}.

The output is a set of data cubes of flux density, $F(x,y,\lambda)$,
where $(x,y)$ is the projection on the sky in arcseconds for a distance
to Taurus of 140\,pc, and the wavelength $\lambda$ is chosen to
cover the three lowest rotational transitions of CO, \13CO, and \C18O,
In principle, there is a tremendous amount of information available
in these cubes for detailed modeling of resolved observations.
However, because our \13CO\ and \C18O\ observations do not have
sufficiently high signal-to-noise ratios to study their spatial distribution,
we spatially and spectrally integrated the model data cubes
and use the line luminosities,
\begin{equation}
L = 4\pi d^2 \int F\,dx\,dy\,dv,
\end{equation}
where $dv=cd\lambda/\lambda$ and $L$ has units Jy\,\kms\ pc$^2$,
to compare with the observations.

\begin{figure}[tb]
\centering
\includegraphics[width=3.4in]{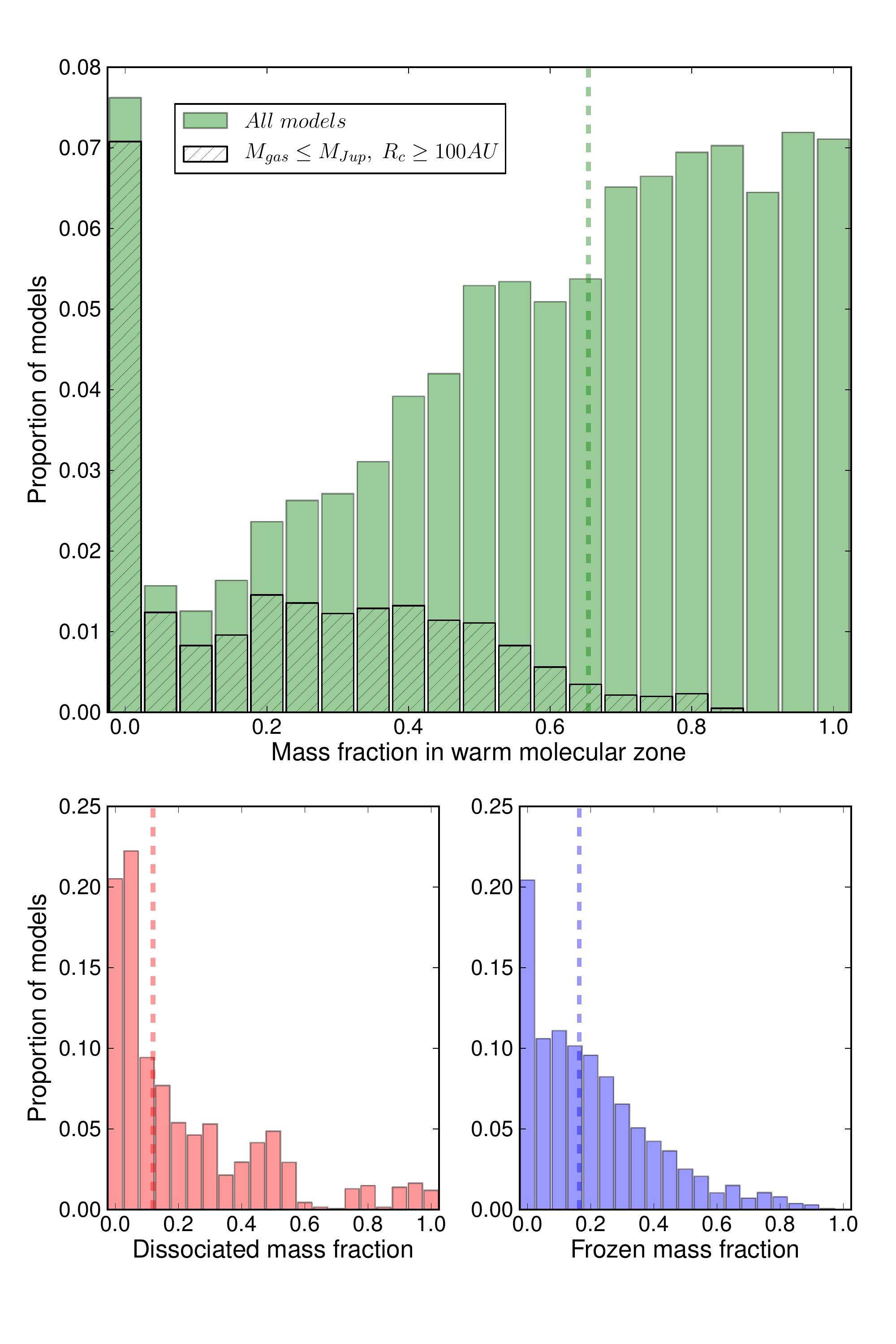}
\caption{
The distribution of gas mass fraction in different regions of the disk
for all the models in the grid. The top panel shows the mass fraction in
the warm molecular layer, which is shielded from dissociation
and is warm enough for CO to be in the gas phase.
The dashed line shows the median of 65\%.
The hashed region shows that most of the models with low CO
fractions are large, low mass disks.
The lower panels show the distribution of the mass fraction
in all models where CO is dissociated,
$N({\rm H}_2) < 1.3\times 10^{21}$\,cm\ee\ (red histogram)
and where it is frozen out, $T < 20$\,K (blue histogram).
The dashed line in each panel indicates the median values,
12\% and 16\% respectively.
}
\label{fig:depletion}
\end{figure}

\begin{deluxetable*}{cc}
\tablecolumns{2}
\tablewidth{0pt}
\tablecaption{Parameter values of the model grid \label{tab:grid}}
\tablehead{
\colhead{Parameter} & \colhead{Range}
}
\startdata
$M_{\rm star}$   & $0.5, 1.0\,M_\odot$ \\
$M_{\rm gas}$    & $10^{-4},3\times 10^{-4},10^{-3},3\times 10^{-3},10^{-2},3\times 10^{-2},10^{-2}~M_\odot$ \\
$r_c$            & 30, 60, 100, 200 AU \\
$\gamma$         & 0.0, 0.75, 1.5 \\
$T_{\rm mid,1}$  & 100, 200, 300 K \\
$T_{\rm atm,1}$  & 500, 750, 1000, 1500 K \\
$q$              & 0.45, 0.55, 0.65 \\ \relax
inclination      & $0^\circ, 45^\circ, 90^\circ$ \\
{[}CO]/[\C18O]   & 550, 1650 \\[-2mm]
\enddata
\end{deluxetable*}

\subsection{Model grid}
\label{sec:grid}
There are a total of nine parameters for each model, listed in the first
column of Table~\ref{tab:grid}.  To explore the dependencies of the line
intensities, we produced a grid of 18144 models that step through the set
of values for each parameter listed in the second column.
The stellar masses bracket those of the Taurus KM stars in our survey.
The disk masses cover a wide range of gas masses from sub-Saturnian
to above the minimum mass required to form the Solar System
\citep{1977MNRAS.180...57W}.
The range for the surface density and temperature profiles are
based on the discussion in
\S\ref{sec:density_structure} and \S\ref{sec:temperature_structure}
and illustrated in Figures~\ref{fig:Sigma} and~\ref{fig:Tgas} respectively.
The three values for the viewing geometry (inclination) span the extremes
of edge-on and face-on, with one intermediate.

We first calculated the mass fraction in those regions of the disk
where CO is dissociated, $f_{\rm dissoc}$,
and where it is frozen out, $f_{\rm freeze}$.
These may overlap in very cold disks with low column densities,
but in general there is a warm molecular layer where CO is
in the gas phase with mass fraction,
$f_{\rm CO} = {\rm max}\{0,1-(f_{\rm dissoc}+f_{\rm freeze})\}$.
Histograms of these fractions over the parameters in the grid are
plotted in Figure~\ref{fig:depletion}.
The median values for all disks are
$f_{\rm dissoc} = 12$\%, $f_{\rm freeze} = 16$\%,
and $f_{\rm CO} = 65$\%.
The dissociation and freeze-out fraction both increase with the
characteristic disk radius, $r_c$, as the outer regions are colder
and of lower density.
The freeze-out fraction is independent of disk mass as the temperature
profile is specified independently and the disk mass only affects the
scaling of the density but not its functional form.
Lower mass disks have lower column densities, however, and therefore
higher dissociation fractions. This can be a dominant factor for the
very lowest mass disks but CO traces the bulk
of the gas mass, $f_{\rm CO} > 50$\%, in 85\% of disks
with $M > M_{\rm Jup}$ for the models in our grid.
These general results demonstrate the feasibility of using CO
isotopologues for constraining gas masses,
at least for disks with the capacity to form giant planets.

\begin{figure*}[!ht]
\centering
\includegraphics[width=3.5in]{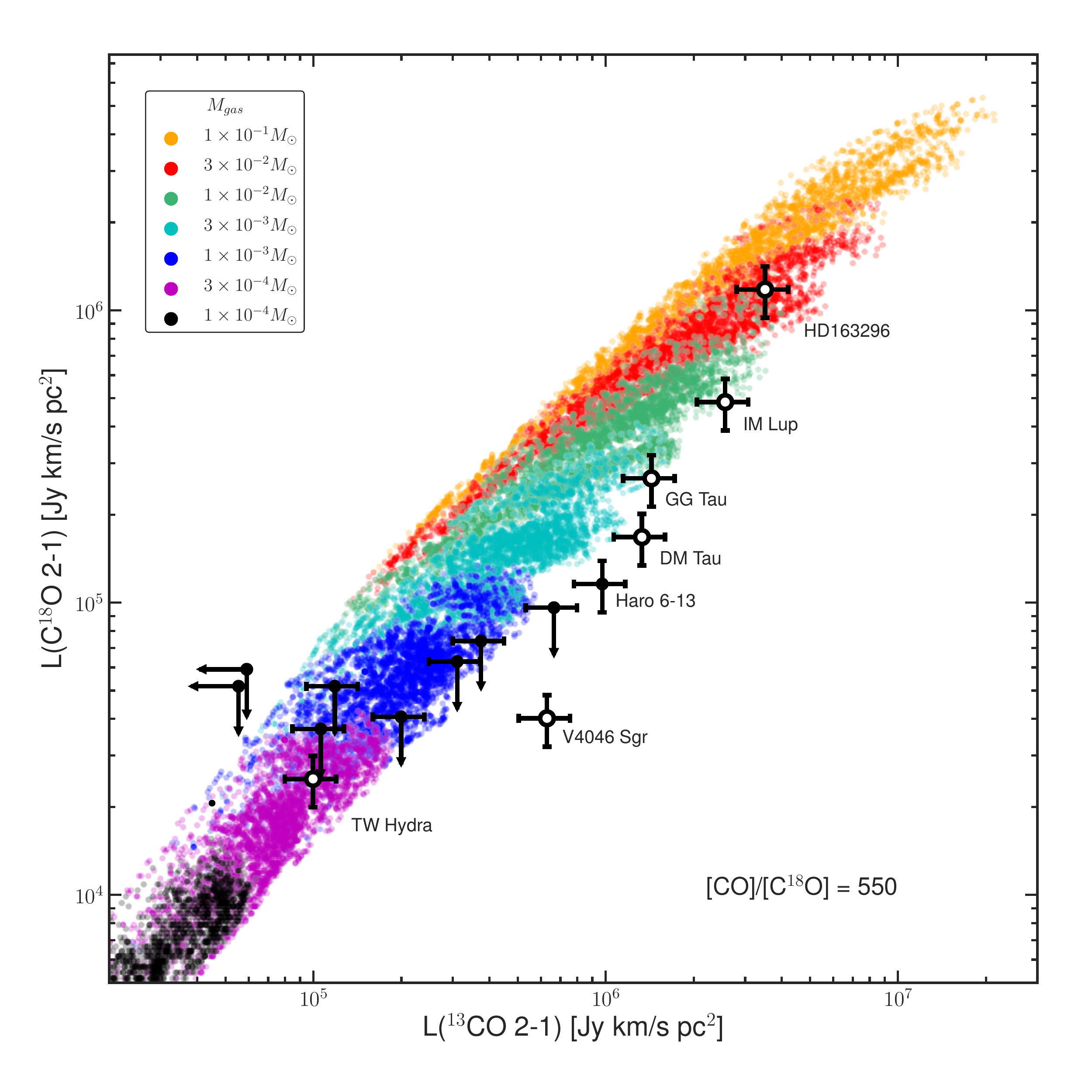}
\includegraphics[width=3.5in]{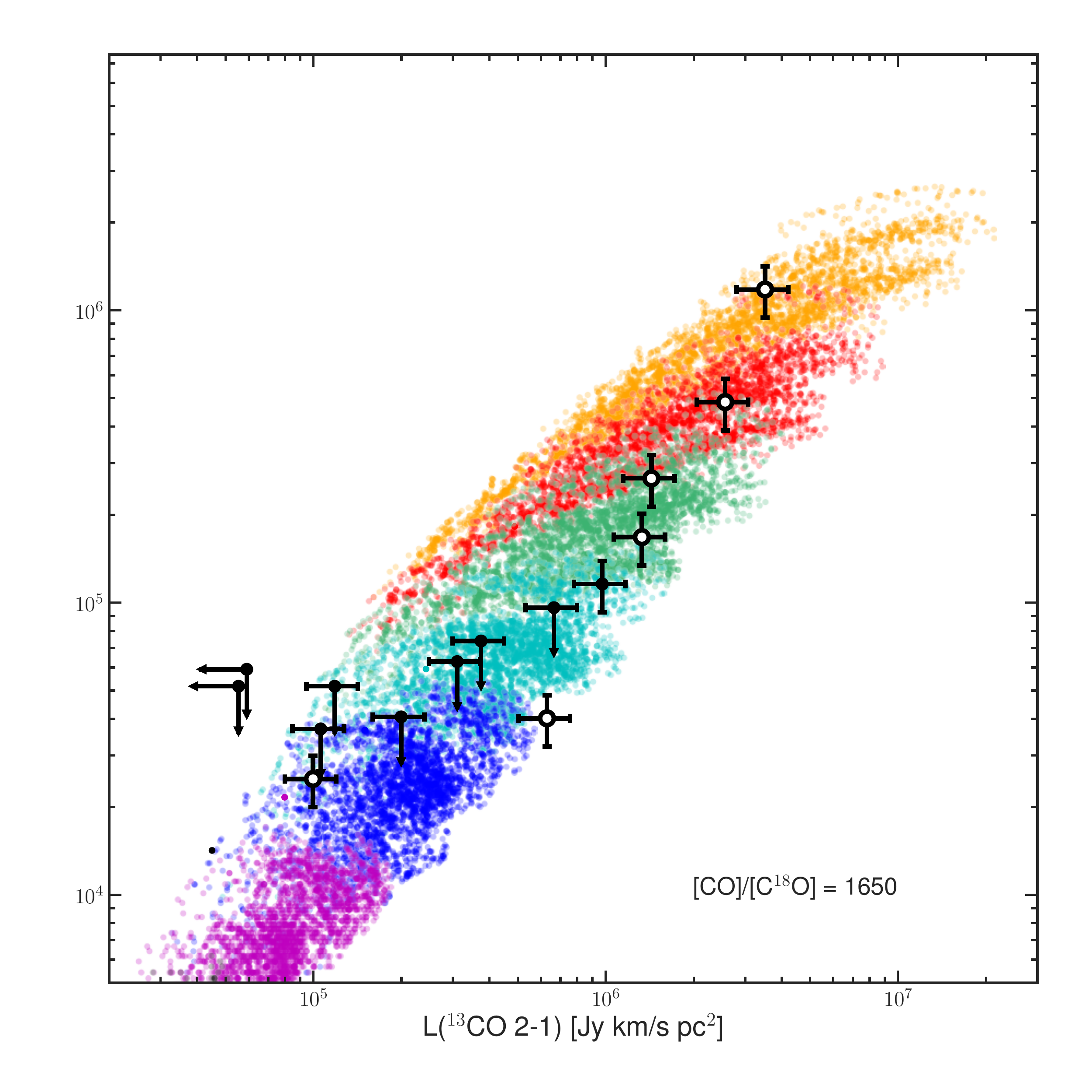}
\caption{
Flux densities for the $J=2-1$ transitions of \13CO\ and \C18O\
for the 18144 calculations in the model grid, color-coded by
gas mass. There is a general trend of
increasing flux density with increasing mass, but a given
flux density of a single line can correspond to a very wide range
of gas masses, about two orders of magnitude for \13CO,
and an order of magnitude for \C18O.  The combination of both
lines, however, allow a much more precise estimate of the gas
mass, typically to within the factor of 3 model grid binning.
The locations of the nine observed disks are noted in black
and the labeled comparison disks in red.
}
\label{fig:twolines}
\end{figure*}

For each model disk, we calculate the line radiative transfer for
CO, \13CO, and \C18O\ at ISM abundances and for \C18O\ at three times
lower abundance to allow for selective photodissociation as discussed above.
The line intensities of the different isotopologues generally follow
each other but there is a large dispersion with CO on account of its
very large optical depth. Furthermore, there are observational problems
with using CO lines as cloud contamination can be severe (and indeed is
directly seen in the SMA data for AA\,Tau, CI\,Tau, DO\,Tau and Haro\,6-13).
We therefore focus on the two isotopologues.
Figure~\ref{fig:twolines} plots the \13CO\ and \C18O\ 2--1 line luminosities
for all the models.  The two lines strongly correlate with each other,
with a median ratio, $L({\rm ^{13}CO~2-1})/L({\rm C^{18}O~2-1})=3.2, 7.0$,
for [CO]/[\C18O] $= 550, 1650$ respectively.
As these median line ratios are only about half the abundance ratios,
we infer that the optical depth in the \13CO\ line is significant.
Nevertheless, disks with different masses tend to lie in different
regions of this diagram. For any given mass, the dispersion in the
line intensities is due partly to CO freeze-out and dissociation
as well as excitation and optical depth.

Further examination of this plot shows that the line intensities are
relatively insensitive to the disk inclination. This is because even
edge-on disks present a large projected area due to their highly flared
vertical structure \citep{1993ApJ...402..280B}.
For a given disk mass, the main factors that affect the line intensities
are the temperature parameters, $T_{\rm atm,1}$ and $T_{\rm mid,1}$.
Disks with high atmospheric temperatures have higher \13CO\ to \C18O\
line ratios, i.e., they lie in the lower part of the envelope
in Figure~\ref{fig:twolines}.
This is likely due to a larger scale height and less saturation of the
the \13CO\ line.
The CO freeze-out fraction is higher in disks with cool midplanes and
the intensities in both lines are correspondingly lower.
The disk size, parameterized mainly by the characteristic radius, $r_c$, 
also broadens the dispersion with small disks tending to have lower
intensities and lower \13CO\ to \C18O\ line ratios. This is also
likely an optical depth effect. However, low mass disks with large $r_c$
may also have low line intensities due to a high CO dissociation fraction.

The variation of the \C18O\ abundance produces a corresponding change
in the \C18O\ line luminosity. There is also a greater dispersion in
the models for any given mass for the lower \C18O\ abundance, however,
which suggests that the \C18O\ emission from the disks with ISM abundances
has a significantly optically thick component.
Nevertheless, the change between the left and right panels of Figure~\ref{fig:twolines}
is almost orthogonal to the variation with mass so the impact on our ability
to constrain gas masses is smaller than we might initially expect.

The combination of all these factors means that there are models
that have the same line luminosity for disks that differ by as much as
a factor of 30 in mass.
Consequently, it is not possible to reliably measure the gas mass from
the luminosity of a single line.
However, the overlap in the two-dimensional scatter plot is much
smaller, about a factor of 10 at most and typically one mass bin
or a factor of 3. The combination of \13CO\ and \C18O\
is therefore a much better diagnostic of gas mass.
The observations are overplotted on the Figure and discussed below.

The driving philosophy behind the parametric model is to create a simple
look-up table for comparison with basic observables in a way that
allows quick, uniform analyses of large samples.
The integrated intensities for the $J=3-2, 2-1, 1-0$ transitions
of CO, \13CO, and \C18O\ (for both relative abundances)
are tabulated in Table~\ref{tab:big_gas_grid},
and available online for the 18144 models in the grid.

\begin{deluxetable*}{ccccccccccccccccccccccc}
%\rotate
\tablecolumns{22}
%\tabletypesize{\scriptsize}
\tabletypesize{\tiny}
\tablewidth{0pt}
\setlength{\tabcolsep}{0.02in} 
\tablecaption{Parametric Model Output\tablenotemark{a} \label{tab:big_gas_grid}}
\tablehead{
\colhead{$M_{\rm star}$} & \colhead{$M_{\rm gas}$} & \colhead{$\gamma$} & \colhead{$R_{\rm c}$} &
\colhead{$T_{\rm mid,1}$} & \colhead{$T_{\rm atm,1}$} &
\colhead{$q$} & \colhead{$i$} & \colhead{$f_{\rm freeze}$} & \colhead{$f_{\rm dissoc}$} &
\colhead{$F_{\rm CO 1-0}$} & \colhead{$F_{\rm CO 2-1}$} & \colhead{$F_{\rm CO 3-2}$} &
\colhead{$F_{\rm ^{13}CO 1-0}$} & \colhead{$F_{\rm ^{13}CO 2-1}$} & \colhead{$F_{\rm ^{13}CO 3-2}$} &
\colhead{$F_{\rm C^{18}O 1-0}$} & \colhead{$F_{\rm C^{18}O 2-1}$} & \colhead{$F_{\rm CO^{18} 3-2}$} &
\colhead{$F_{\rm C^{18}O 1-0}^{\rm low}$} & \colhead{$F_{\rm C^{18}O 2-1}^{\rm low}$} & \colhead{$F_{\rm CO^{18} 3-2}^{\rm low}$} \\
\colhead{($M_\odot$)} & \colhead{($M_\odot$)} & \colhead{} & \colhead{(AU)} &
\colhead{(K)} & \colhead{(K)} &
\colhead{} & \colhead{($^\circ$)} & \colhead{} & \colhead{} &
\colhead{(Jy\,\kms)} & \colhead{(Jy\,\kms)} & \colhead{(Jy\,\kms)} &
\colhead{(Jy\,\kms)} & \colhead{(Jy\,\kms)} & \colhead{(Jy\,\kms)} &
\colhead{(Jy\,\kms)} & \colhead{(Jy\,\kms)} & \colhead{(Jy\,\kms)} &
\colhead{(Jy\,\kms)} & \colhead{(Jy\,\kms)} & \colhead{(Jy\,\kms)}
}
\startdata
0.5 & 1.00E-04	& 0 & 30 & 100 & 500 & 0.45 &  0 & 0.039 & 0.164 & 0.150 & 0.841 & 2.126 & 0.022 & 0.142 & 0.390 & 0.004 & 0.043 & 0.125 & 0.002 & 0.017 & 0.053 \\
0.5 & 1.00E-04	& 0 & 30 & 100 & 500 & 0.45 & 45 & 0.039 & 0.164 & 0.158 & 0.835 & 2.043 & 0.023 & 0.162 & 0.452 & 0.005 & 0.045 & 0.133 & 0.002 & 0.018 & 0.055 \\
0.5 & 1.00E-04	& 0 & 30 & 100 & 500 & 0.45 & 90 & 0.039 & 0.164 & 0.092 & 0.391 & 0.862 & 0.020 & 0.136 & 0.373 & 0.004 & 0.041 & 0.121 & 0.002 & 0.017 & 0.052 \\
0.5 & 1.00E-04	& 0 & 30 & 100 & 750 & 0.45 &  0 & 0.029 & 0.169 & 0.212 & 1.407 & 3.862 & 0.020 & 0.152 & 0.455 & 0.004 & 0.039 & 0.122 & 0.001 & 0.015 & 0.048 \\
0.5 & 1.00E-04	& 0 & 30 & 100 & 750 & 0.45 & 45 & 0.029 & 0.169 & 0.224 & 1.403 & 3.721 & 0.021 & 0.167 & 0.511 & 0.004 & 0.040 & 0.127 & 0.001 & 0.015 & 0.050 \\[-2mm]
\enddata
\tablenotetext{a}{This table is published in its entirety in the electronic
edition. A portion is shown here for guidance regarding its form and content.}
\end{deluxetable*}

\section{Results}
\label{sec:results}
The utility of the model grid is best demonstrated through application.
Observations of \13CO\ and \C18O\ 2--1 integrated line intensities
for six well studied disks from the literature and for the sample of
nine less well known but perhaps more typical disks from our SMA Taurus
survey are overlaid on Figure~\ref{fig:twolines}.
When detected, the measurements generally have a signal-to-noise
ratio greater than 5 and the calibration uncertainty is greater than
the measurement error.
Aside from the numerous \C18O\ upper limits in the case of the survey disks,
therefore, we have placed error bars of $\pm 20$\% on the data.

\subsection{Comparison with well studied disks}
\label{sec:results_famous_disks}
A handful of disks have received more attention in the literature than
others at millimeter wavelengths primarily due to their apparent brightness.
That is, they are either relatively close or relatively massive.
We found six disks with published results on the 2--1 lines of
\13CO\ and \C18O\ in the literature and compiled their integrated
intensities in Table~\ref{tab:famous_disks}.

\begin{deluxetable*}{lcccccccccc}
\tablecolumns{11}
\tablewidth{0pt}
\tablecaption{Comparison disk 1.3\,mm fluxes\label{tab:famous_disks}}
\tablehead{
\colhead{} &
\multicolumn{2}{c}{$F_{\rm cont}$ (mJy)} &
\multicolumn{2}{c}{$F_{\rm CO}$ (Jy\,\kms)} &
\multicolumn{2}{c}{$F_{\rm ^{13}CO}$ (Jy\,\kms)} &
\multicolumn{2}{c}{$F_{\rm C^{18}O}$ (Jy\,\kms)} &
\colhead{Dist.} &
\colhead{} \\
\colhead{Source} &
\colhead{Value} & \colhead{$\sigma$} &
\colhead{Value} & \colhead{$\sigma$} &
\colhead{Value} & \colhead{$\sigma$} &
\colhead{Value} & \colhead{$\sigma$} &
\colhead{(pc)} &
\colhead{Ref.}
}
\startdata
TW Hydra  &  540 & 30  & 17.5  & 1.8  &  2.72 & 0.18 &  0.68 & 0.18 &   54  & 1,2,3 \\
V4046 Sgr &  283 & 28  & 34.5  & 3.5  &  9.4  & 0.9  &  0.6  & \nodata &   73  & 4 \\
DM Tau    &  109 & 13  & 14.9  & 0.4  &  5.40 & 0.13 &  0.68 & 0.07 &  140  & 6,7 \\
GG Tau    &  593 & 53  & 21.5  & 1.6  &  5.82 & 0.19 &  1.08 & 0.10 &  140  & 6,7 \\
IM Lup    &  195 & \nodata & 22.3  & \nodata &  5.65 & \nodata &  1.07 & \nodata &  190  & 8 \\
HD 163296 &  670 & 0.7 & 54.17 & 0.39 & 18.76 & 0.24 &  6.30 & 0.16 &  122  & 5 \\[-6mm]
\enddata
\tablenotetext{~}{
References:
1--\citet{2006ApJ...636L.157Q},
2--\citet{2012ApJ...757..129R},
3--\citet{2013Sci...341..630Q},
4--\citet{2013ApJ...775..136R},
5--\citet{2011ApJ...740...84Q},
6--\citet{2005ApJ...631.1134A},
7--\citet{1997A&A...317L..55D},
8--\citet{2009A&A...501..269P}
}\\
\end{deluxetable*}

With the exception of the Herbig Ae star, HD\,163296, the stars
in this sample have similar masses and luminosities as the low
mass Taurus stars in our SMA survey.
The range of temperatures in the model grid should therefore be 
appropriate for these systems.
However, several of these disks are known to have
inner cavities. This is particularly large in the case
of the GG\,Tau binary system \citep[$R\simeq 180$\,AU,][]{1994A&A...286..149D},
but also resolved in TW\,Hydra
\citep[$R\simeq 3$\,AU,][]{2014A&A...564A..93M, 2007ApJ...664..536H},
DM\,Tau \citep[$R\simeq 19$\,AU,][]{2011ApJ...732...42A},
and in the spectroscopic binary V4046\,Sgr
\citep[$R\simeq 29$\,AU,][]{2013ApJ...775..136R}.
The model grid does not incorporate large inner holes but
since most of the mass lies at large radii for surface
density profiles that are flatter than $\Sigma(r)\propto r^{-2}$
and because we are interested in low lying CO rotational lines
that are readily excited in the outer parts,
we expect the grid to be broadly applicable to these observations.
Indeed we find that the \13CO\ and \C18O\ line luminosities,
while spanning a range of about a factor of 30,
lie within our model results. The models with the reduced
\C18O\ abundance fit the data better in general which suggests that
selective photodissociation is an important effect to consider
in interpreting observations of this species.

\begin{figure}[tb]
\centering
\includegraphics[width=3.5in]{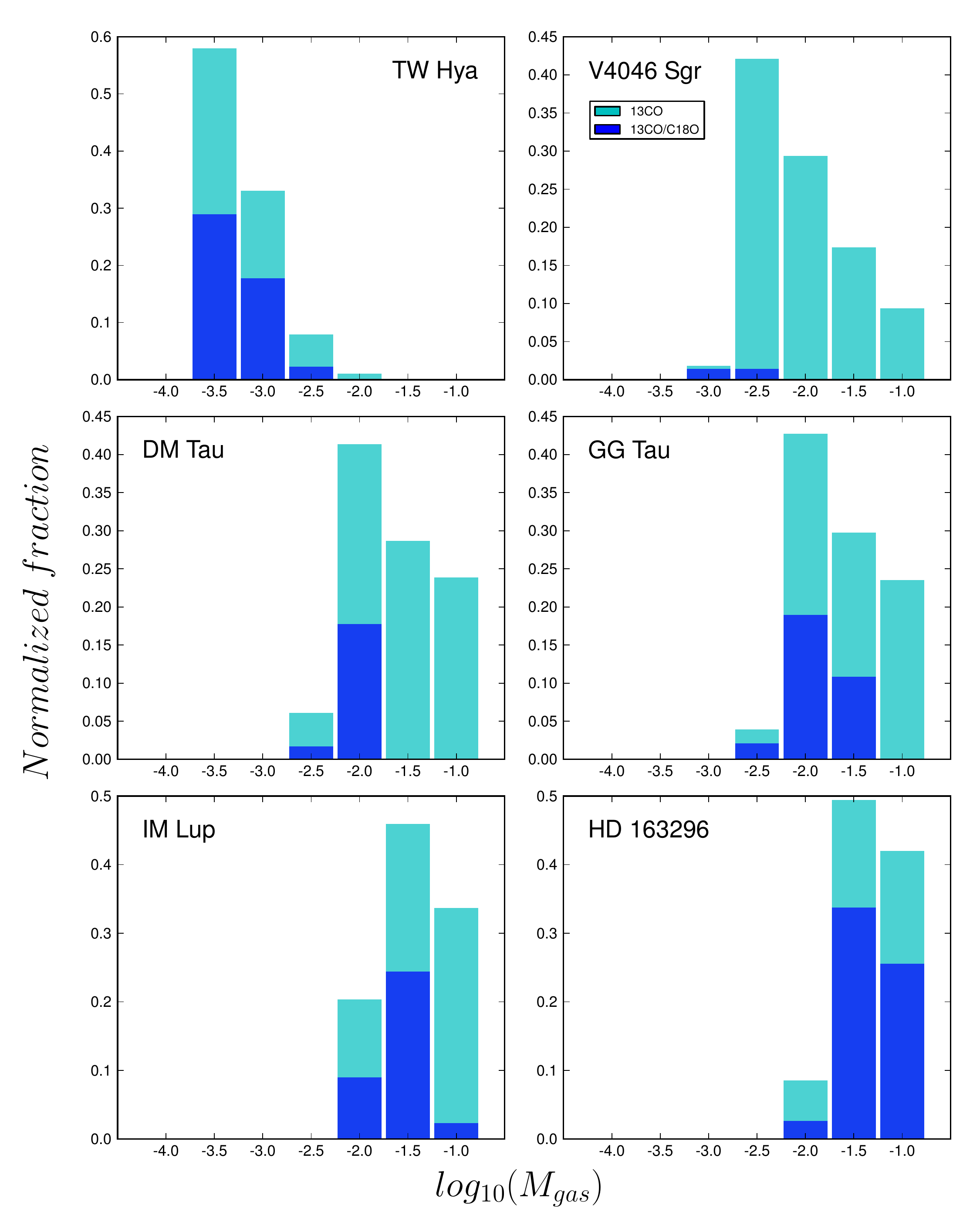}
\caption{
The results of model grid fitting for gas masses
for the six comparison disks listed in Table~\ref{tab:famous_disks}.
Each panel is a histogram of the proportion of the
models that match the observed line luminosities.
The cyan histogram shows fits to the \13CO\ 2-1 line luminosity only
and the blue is a subset that fits both the \13CO\ and \C18O\
line luminosities.
For each source, the histograms are normalized such that the cyan
histogram sums to a total fraction of 1.
}
\label{fig:Mgas_fits_famous}
\end{figure}

Histograms of the gas masses that fit the \13CO\ line alone and the
combination of \13CO\ and \C18O\ (for any other model parameter)
are shown in Figure~\ref{fig:Mgas_fits_famous}.
The inferred values range from below a Jupiter mass in the
case of TW\,Hydra to above the MMSN for most others.
The low mass inferred for the TW\,Hydra disk is consistent
with detailed modeling of Herschel observations and CO and \13CO\ 3--2
lines \citep{2010A&A...518L.125T}
but notably discrepant with the HD mass measurement
by \citet{2013Natur.493..644B}.
We discuss potential implications of this in \S\ref{sec:discussion}.
\citet{1997A&A...317L..55D} estimate masses for DM\,Tau and GG\,Tau
based on modeling their chemistry that are consistent with our values.
The Herbig Ae star, HD\,163296 is the most massive and luminous star
in the sample and has the most massive disk. It lies near the boundaries
of our model grid parameters, which is designed to model T Tauri stars.
Nevertheless, our inferred gas mass lies within a factor of two of
the detailed modeling of multiple resolved CO and isotopologue lines by
\citet{2011ApJ...740...84Q}.

There is no independent gas mass determination for IM\,Lup so
we cannot compare with our results in this case.
The only major discrepancy between our model grid results and detailed
modeling of the CO lines in individual sources is V4046\,Sgr
where our estimated gas mass is significantly lower than a
three-component model by \citet{2013ApJ...775..136R}.
We note, however, that the \C18O\ line was only marginally detected
in that study and was not used to constrain their fits.
The disk location in Figure~\ref{fig:twolines} shows it to have an
abnormally low \C18O\ to \13CO\ line luminosity relative
to other disks and very few models in our grid match both lines.

\subsection{Fits to the surveyed disks}
\label{sec:results_survey_disks}
\subsubsection{Case study: Haro 6-13}
\label{sec:Haro6-13}
\begin{figure}
\centering
\includegraphics[width=3.5in]{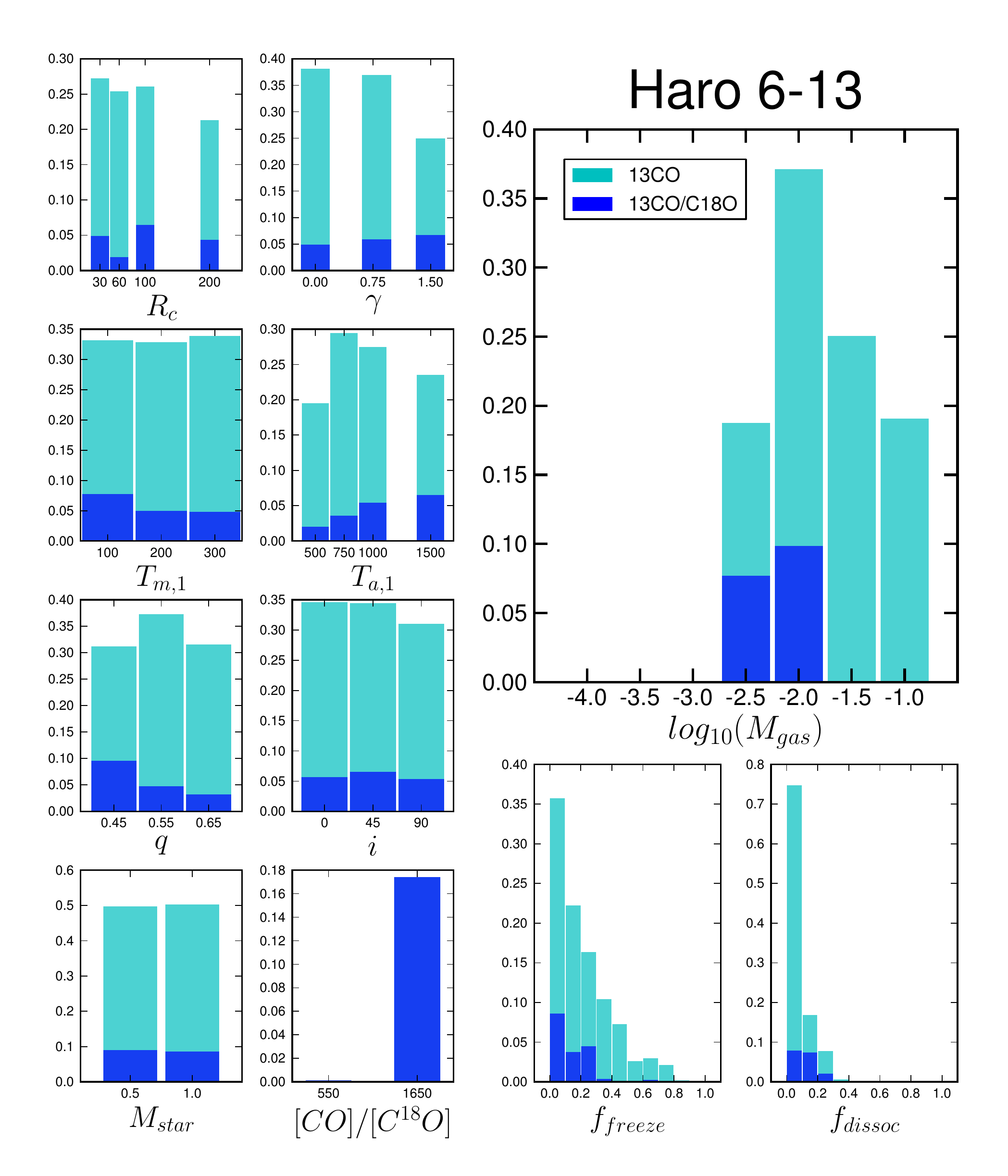}
\caption{
Histograms of the nine parameters and the derived dissociated and frozen
mass fractions for the models that match the Haro\,6-13 observations to
within 20\%.
The color coding and
normalization is the same as in Figure~\ref{fig:Mgas_fits_famous}.
}
\label{fig:Haro6-13}
\end{figure}

Haro\,6-13 was the only disk in our SMA survey that was detected in \C18O\
and therefore more tightly constrains the model parameters than the others.
Figure~\ref{fig:Haro6-13} plots histograms of each of the nine
parameters for fits to \13CO\ alone and for \13CO\ and \C18O.

Considering the grid matches to the \13CO\ line alone shows a wide
range in gas mass and all other parameters.
This reiterates the expectation from from Figure~\ref{fig:twolines}
that a single line does not provide a strong constraint. Disks with gas
masses that range over one order of magnitude match the line
luminosity and the distribution of $f_{\rm freeze}$ is widely
spreaad. As a moderately massive disk, however, the dissociation fraction
$f_{\rm dissoc}$ is fairly small.

Fits to both the \13CO\ and \C18O\ lines provide stronger constraints
on the gas mass, to within a factor of 3 and somewhat less than the MMSN.
The relatively weak \C18O\ emission indicates a high [CO]/[\C18O]
ratio as expected for selective photodissociation.
The histograms of the disk structure and temperature parameters
do not show any strong preferences for particular values,
which is simply a reflection that the gas mass is the primary
determinant of the line intensities.
This relative insensitivity means that unknowns such as the
precise vertical temperature distribution are not strongly limiting factors
in our ability to determine disk masses. A further important implication
is the feasibility of this modeling approach for quantifying large surveys
without specific detailed modeling of each individual system.

\subsubsection{The full survey}
\label{sec:results_all_disks}
We have carried out the same exercise as above for the other
eight disks in the SMA survey.
Although \C18O\ was not detected in any of these objects, we obtained
sensitive upper limits that prove critical for constraining
the gas masses, especially in the case of those sources with
reasonably strong \13CO\ emission.

\begin{figure}
\centering
\includegraphics[width=3.5in]{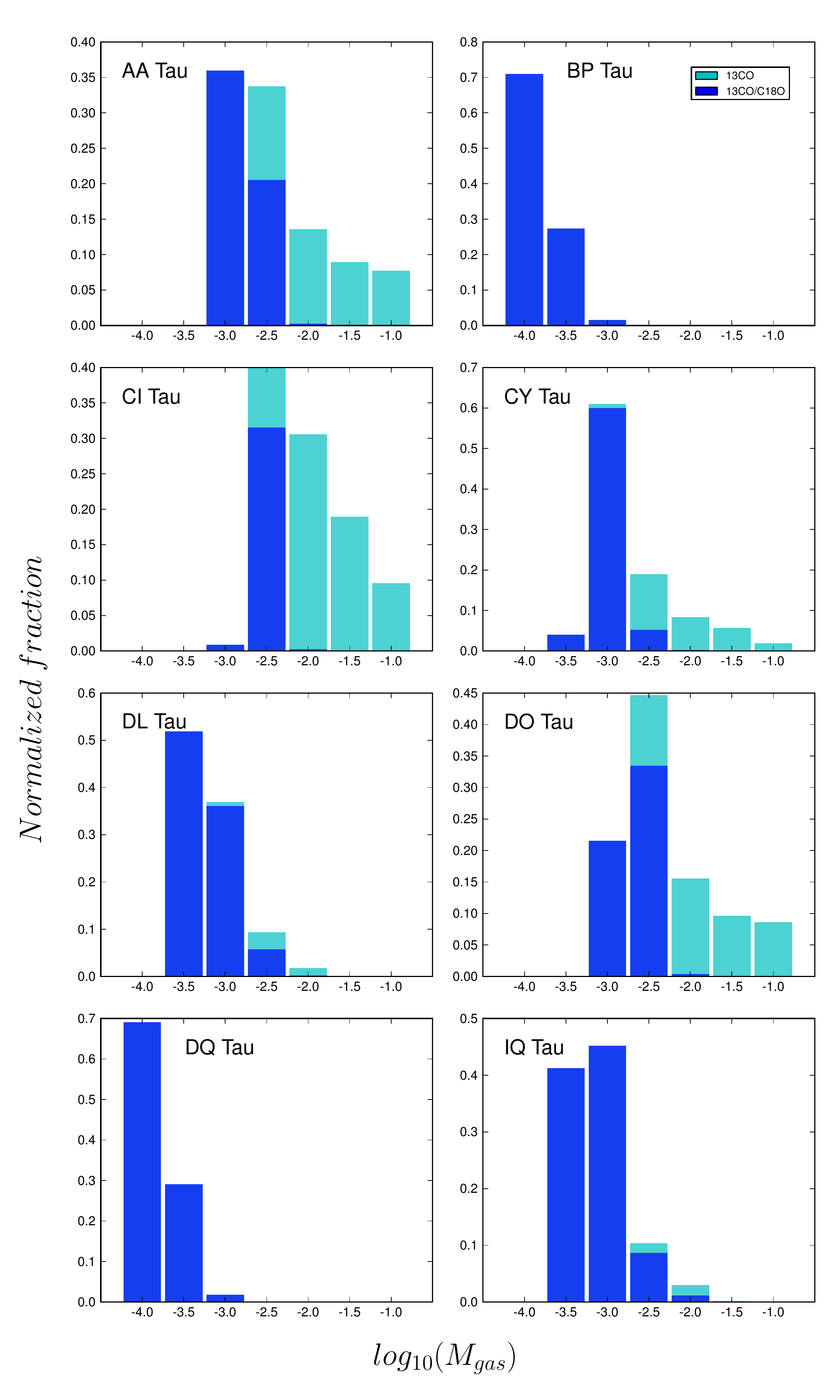}
\caption{
The results of model grid fitting for gas masses
for the eight other disks in the SMA survey.
Each panel is a histogram of the proportion of the
models that match the observed line luminosities. The color coding and
normalization is the same as in Figure~\ref{fig:Mgas_fits_famous}.
}
\label{fig:Mgas_fits_survey}
\end{figure}

Figure~\ref{fig:Mgas_fits_survey} plots the histograms of the gas mass
for the fits to the same line combinations as above.
Models that range over a factor of 30 or more can match
the \13CO\ line alone.
The additional constraint from the \C18O\ limit pins down
down the mass range to about a factor of 3 for 
AA\,Tau, CI\,Tau, CY\,Tau, and DO\,Tau,
and forces them to the lower range of the \13CO\ fits.
This indicates that these disks are reasonably warm
and have sufficient flaring that \13CO\ optical depth effects
are not hiding a lot of mass.
The inferred masses are low, $M_{\rm gas} \simeq 1-3\,M_{\rm Jup}$.

The information from the \C18O\ line does not have the
same leverage for the weaker sources DL\,Tau and IQ\,Tau
and is inconsequential for the case of BP\,Tau and DQ\,Tau
which were not detected even in \13CO\ (but the presence
of CO demonstrates that some molecular gas exists).
We conclude that these disks all have very low masses,
$M_{\rm gas} \simlt M_{\rm Jup}$.
As with the histograms in Figure~\ref{fig:Haro6-13} for Haro\,6-13,
we find for all the disks that the low excitation CO isotopologue
lines do not strongly constrain the other model parameters.

The primary result from this work is that the 9 Taurus disks in our
survey all have low gas masses, well below the MMSN, and 5 have
masses no greater than a Jupiter mass.
This is surprising in that the masses are much lower than expected
from the dust emission and implies low gas-to-dust ratios.
It also has clear implications for planet formation and/or
disk chemistry that are discussed in \S\ref{sec:discussion}.

\subsection{Dust masses and gas-to-dust ratios}
\label{sec:dust_masses}
For comparison with the gas masses from the model fits above,
we calculated the dust masses using the standard equation,
\begin{equation}
M_{\rm dust} = \frac{F_{\rm cont} d^2}{\kappa_\nu B_\nu(T_{\rm dust})},
\end{equation}
assuming optically thin emission and a constant temperature, $T_{\rm dust}$.
Here $F_{\rm cont}$ is the continuum flux density listed in
Tables~\ref{tab:fluxes} and \ref{tab:famous_disks},
$d$ is the distance,
$B_\nu$ is the Planck function at the observing frequency, $\nu$,
and $\kappa_\nu = 10(\nu/1000\,{\rm GHz}){\rm cm}^{2}\,{\rm g}^{-1}$
is the dust grain opacity from the prescription in
\citet{1991ApJ...381..250B} without the implicit gas-to-dust ratio.
Following \citet{2013ApJ...771..129A}
we scale the dust temperature with the stellar luminosity,
$T_{\rm dust} = 25 (L_{\rm star}/L_\odot)^{1/4}$\,K.
There are the well known caveats regarding the possibility of
optically thick emission in the inner regions of the disk and
missing mass in large grains \citep[see, e.g.,][]{2011ARA&A..49...67W},
and this measure of the dust mass is therefore a lower limit
to the total mass of solids.

\begin{figure}
\centering
\includegraphics[width=3.5in]{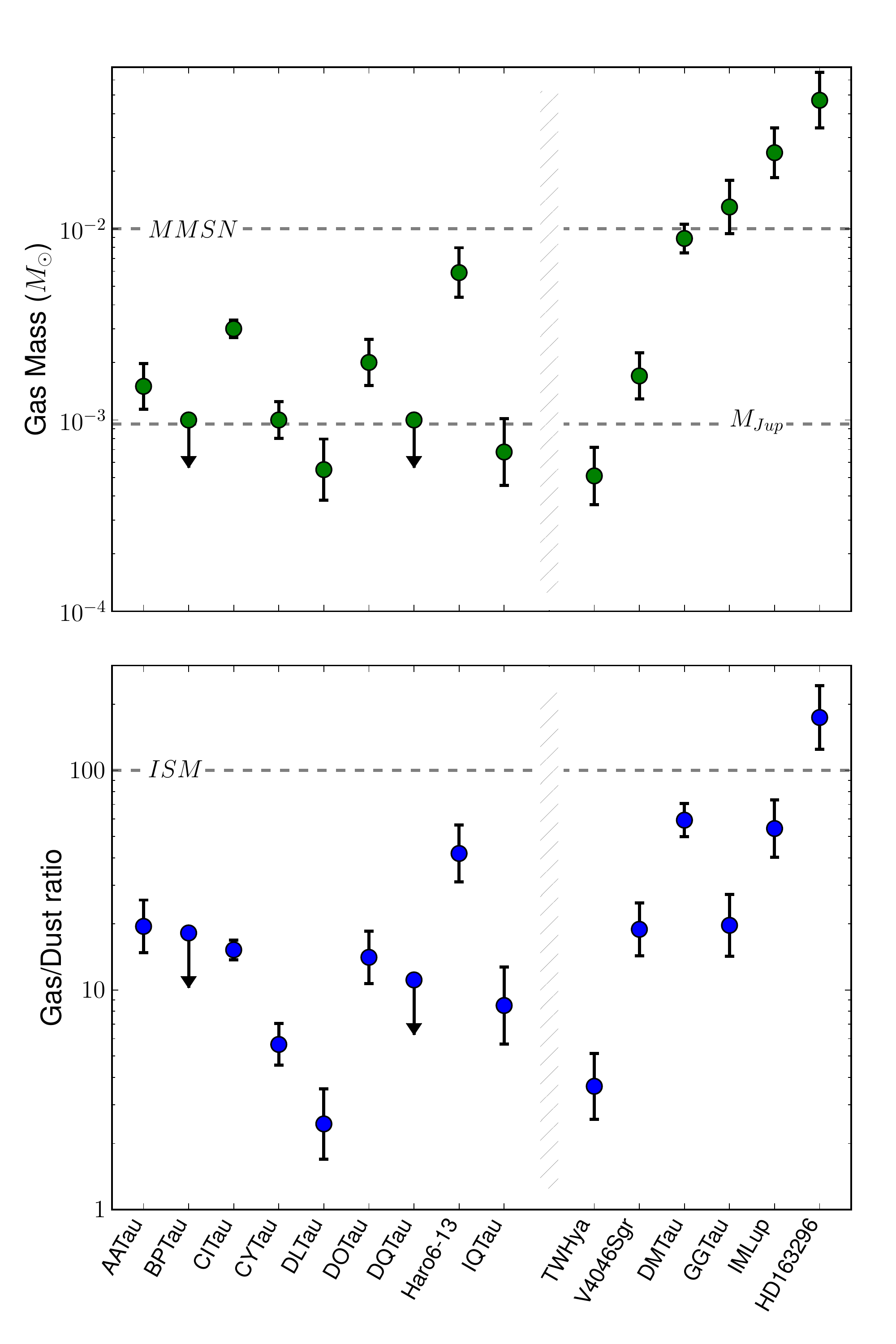}
\caption{
The fitted gas masses and inferred gas-to-dust ratios for the nine
Taurus disks in the SMA survey and the six comparison disks.
The upper panel shows the mean gas masses
with dashed lines at the Minimum Mass Solar Nebula ($0.01\,M_\odot$)
and a Jupiter mass for comparison.
Uncertainties are estimated from the range of the fits and the factor
of 3 mass binning,
except for the two sources, BP Tau and DQ Tau, which were undetected
in \13CO\ and for which we can only determine upper limits.
The lower panel is the ratio of gas mass to the dust mass derived from
the continuum flux density, and is compared to the ISM value of 100
shown by the dashed line.
The vertical hashes in each panel divide the SMA survey from
the comparison disks as these are a more heterogeneous group in terms
of stellar type and disk structure.
}
\label{fig:masses}
\end{figure}

The derived gas and dust masses, and their ratios, are listed in
Table~\ref{tab:masses} for the SMA survey and comparison disks.
The gas masses here are weighted averages from the histograms in
Figures~\ref{fig:Mgas_fits_famous}, \ref{fig:Haro6-13}, \ref{fig:Mgas_fits_survey}.
The range in dust masses is smaller than the range in gas masses
which demonstrates that there is a significant dispersion in the
gas-to-dust ratio.  Figure~\ref{fig:masses} plots the gas masses
with markers at the MMSN and Jupiter, and the gas-to-dust ratios with the
fiducial ISM value of 100.

All the disks in the SMA survey and the two nearby comparison disks,
TW\,Hydra and V4046\,Sgr, have gas masses below the MMSN and
gas-to-dust ratios below 100. Several have extremely low gas
masses, comparable or below that of Jupiter, and all but Haro\,6-13
have gas-to-dust ratios below 30.
The comparison sample is biased toward bright objects and the
more distant members, DM\,Tau, GG\,Tau, IM\,Lup, and HD\,163296
are intrinsically massive. Their inferred gas masses are at the MMSN
level or higher. Nevertheless, with the exception of HD\,163296,
these disks also have gas-to-dust ratios that are lower than 100.

The second main result from this work is that the gas-to-dust ratio
is both low and varies widely from disk to disk.
The mean ratio for the disks in the SMA survey is 16 with a standard
deviation of 11. Including the comparison disks in addition,
the mean is 25 with a standard deviation of 24.
It appears that the composition of these Class II disks has
evolved significantly from their initial conditions.

\begin{deluxetable}{lccr}
\tablecolumns{4}
\tablewidth{0pt}
\tablecaption{Gas and dust masses \label{tab:masses}}
\tablehead{
\colhead{Source} &
\colhead{$M_{\rm gas} (10^{-4}\,M_\odot)$} &
\colhead{$M_{\rm dust} (10^{-4}\,M_\odot)$} &
\colhead{Ratio}
}
\startdata                                     % mean/sd of log(Mgas)
AA Tau    &   15  & 0.77 &   19  \\            %  -2.82  0.24
BP Tau    & $<10$ & 0.55 & $<18$ \\            %  -3.5   upper limit
CI Tau    &   30  & 2.0  &   15  \\            %  -2.53  0.09
CY Tau    &   10  & 1.8  &    6  \\            %  -2.99  0.19
DL Tau    &    5  & 2.2  &    2  \\            %  -3.26  0.32
DO Tau    &   20  & 1.4  &   14  \\            %  -2.70  0.24
DQ Tau    & $<10$ & 0.90 & $<11$ \\            %  -3.5   upper limit
Haro 6-13 &   60  & 1.4  &   43  \\            %  -2.23  0.26
IQ Tau    &    7  & 0.80 &    9  \\[0.5ex]     %  -3.17  0.35
\hline\\[-2ex]
TW Hydra  &    5  & 1.4  &    4  \\            %  -3.29  0.30
V4046 Sgr &   17  & 0.90 &   19  \\            %  -2.76  0.24
DM Tau    &   90  & 1.5  &   60  \\            %  -2.05  0.15
GG Tau    &  130  & 6.6  &   20  \\            %  -1.87  0.28
IM Lup    &  250  & 4.6  &   53  \\            %  -1.61  0.26
HD 163296 &  470  & 2.7  &  170  \\[-3mm]      %  -1.33  0.29
\enddata
\end{deluxetable}

\section{Discussion}
\label{sec:discussion}
Given the small set of assumptions in our disk gas model and the
wide range of parameters used to create the grid, it is not surprising
that we are able to find model sets that match the observed integrated
line intensities for each of the sources in
Tables~\ref{tab:fluxes} and \ref{tab:famous_disks}.
It is more noteworthy, however, that the
line intensities are more strongly dependent on mass than the
parameters describing the disk density or temperature structure,
at least for the range of expected, or in some cases observed, values.
This then leads to the simple, but powerful, result that we can determine
gas masses from the unresolved integrated emission of two lines
reasonably accurately and independently of the details of their structure.

Our finding of low gas masses and low gas-to-dust ratios
is somewhat unexpected.
Hence it is worth noting that, even without detailed modeling,
the observations here have line-to-continuum ratios $\sim 1-5$
for the \13CO\ 2--1 line, which is much lower than the observed
values, $\simgt 100$ for star forming cores in NGC\,1333
\citep{2006A&A...451..539S, 1998A&A...334..269L}.
In this basic observational sense, therefore, disks lie in
a fundamentally different regime of parameter space.
Part of the difference may be due to grain growth from microns to
millimeters, resulting in
higher emission efficiency at millimeter wavelengths, but most
is from the weaker line emission.  As the transition is readily
excited and the emission is not substantially optically thick,
the low line intensities imply lower column densities.

There have been previous suggestions of low gas-to-dust
ratios in Class II disks.
It was realized early on that the CO emission is relatively weak in
TW Hydra, either because of a low gas mass relative to the dust,
or depletion within the warm molecular layer 
\citep{1997Sci...277...67K,2001A&A...377..566V}.
In addition,
\citet{2003A&A...402.1003D} found weak CO line emission toward BP\,Tau
and, as with our SMA observations, were unable to detect it in \13CO.
They postulated that the gas is disappearing before the dust
during the transition to a non-accreting Class III source.
Subsequent observations by \citet{2013A&A...549A..92G} provided tighter
constraints on the \13CO\ line luminosity and showed no emission
from other molecules.
The non-detection of CQ\,Tau in CI by \citet{2010A&A...520A..61C}
was also interpreted as an indication of a very low gas-to-dust ratio
and shorter depletion timescales for gas relative to dust.
Note that these interpretations of low gas-to-dust ratios
are a global disk average and are not inconsistent with the
existence of locally high values in the outer disk
due to inward migration of dust \citep{2012ApJ...744..162A}.

There are two explanations for our results of low disk gas
masses and low gas-to-dust ratios,
both with important implications for planet formation.
Taken at face value, the preferential loss of gas relative to dust
may be due to grain growth and settling toward the disk mid-plane,
which leaves behind a gas-rich disk atmosphere that
may be lost through photo-evaporation \citep{2013arXiv1311.1819A}
or accreted onto the star \citep{1996ApJ...457..355G}.
As an indication of the plausibility of the latter, we note that
gas accretion rates integrated over protostellar ages are about an
an order of magnitude higher than the disk masses derived from
the dust continuum with a gas-to-dust ratio of 100
\citep{2007ApJ...671.1800A}.
In only about 10\% of the protostellar lifetime, therefore,
most of the disk may be accreted onto the star and, if the
accreted material is gas-rich, the gas-to-dust ratio of the
surviving disk that we observe will be low.

So little gas remains in our surveyed disks, less than a few Jupiter
masses spread over tens of AU, that gas giant planets are unlikely to form.
This is consistent with the low numbers of Jovian planets seen
around stars with masses $\sim 0.6-0.7\,M_\odot$,
either through radial velocity surveys or transits\footnotemark
\footnotetext{based on exploring the Exoplanet database,
http://exoplanets.org},
or direct imaging \citep{2013ApJ...777..160B}.
We would expect higher CO isotopologue line intensities
(though not necessarily higher line-to-continuum ratios)
from disks around more massive stars, $\sim 1\,M_\odot$,
that more typically host Jupiter mass planets.
We should also see higher gas-to-dust ratios in earlier
evolutionary phases before most of the disk accretes or photoevaporates,
although a direct comparison of \13CO\ lines may be complicated
by confusion between disk and envelope in Class 0 and I sources.

Alternatively, we would have under-estimated the gas masses
if our assumption of an ISM-like CO abundance relative to H$_2$,
$x({\rm CO}) = 10^{-4}$, is wrong.
To achieve a gas-to-dust ratio of 100 in the surveyed disks
would require CO abundances lower by an average factor of 6.
This was indeed the suggestion of \citet{2013ApJ...776L..38F}
to reconcile observations of \C18O\ 2--1 in TW\,Hydra
with the high mass, $M_{\rm gas}> 0.05\,M_\odot$,
estimated from Herschel observations of a single HD line
\citep{2013Natur.493..644B}.
They proposed that vertical and radial mixing of gas through the
snowline might remove CO and substantially lower its abundance in
the warm molecular layer.
We note, however, the contrary point made by \citet{2011ApJS..196...25S}
that the CO chemical timescale is shorter than the diffusion timescale
and therefore that its abundance should not be strongly affected by transport.

Our model fits to the TW Hydra \13CO\ and \C18O\ observations
have similar density and temperature profiles as in 
\citet{2013ApJ...776L..38F} and indicate a very low mass,
with a mean value $M_{\rm gas}=5\times 10^{-4}\,M_\odot$
that is 100 times lower than the HD derived mass,
but with a wide range and an upper limit of
$3\times 10^{-3}\,M_\odot$ or a factor of 17 lower.
To lower the CO abundance so radically requires that the midplane
be enhanced in CO ices or related chemical by-products by a factor
of $\sim 1/f_{\rm freeze}$ where $f_{\rm freeze}$ is the disk
mass fraction at temperatures below 20\,K.
The fits to our grid span a wide range $0-0.8$ with mean value 0.3.
The inverse is clearly very poorly constrained.
What might happen to such an icy CO reservoir is unclear;
planet-bearing stars are enriched in carbon and oxygen
\citep{2011ApJ...735...41P},
and the gas in the $\beta$\,Pic planetesimal debris disk is also
carbon-enriched \citep{2006Natur.441..724R},
but the Earth is deficient in carbon relative to the Sun
\citep{2001E&PSL.185...49A},
as are asteroids around white dwarfs \citep{2012ApJ...750...69J}.

Any uncertainty in the CO-to-H$_2$ abundance will propagate into the
disk masses derived by the method described here.
Without other HD observations or other independent mass estimates,
it is unclear whether TW\,Hydra is unusual.
It is thought to be a relatively old disk \citep{2004ARA&A..42..685Z}
so its chemistry may be more advanced than the Taurus disks in our survey
\citep{1997ApJ...486L..51A}.
When allowance is made for its close distance,
Figure~\ref{fig:twolines} demonstrates that it has
much weaker line emission than the rest of the comparison sample,
though it is not such an outlier compared to more typical Class II
disks in our Taurus sample.
It should be possible to learn more about the chemical composition of
the gas in the cold regions of the disk through sensitive observations
of additional molecules in the near future.

The transitions of \13CO\ and \C18O\ have similar frequencies
and can be observed simultaneously for $J=2-1$ with the SMA
and also for $J=1-0$ and $3-2$ with the
Atacama Large Millimeter/Sub-Millimeter Array (ALMA).
Surveys to measure gas masses and gas-to-dust ratios can therefore be
efficiently carried out.
Furthermore, with higher signal-to-noise, resolved images of the line emission,
it should be possible to build on the modeling technique described here
and measure the radial profile of the gas surface density and gas-to-dust ratio
in hundreds of disks.  This will open up new avenues for understanding the origins
of planets and the influence of initial conditions on the tremendous range of
observed exoplanetary systems.

\section{Summary}
\label{sec:summary}
We have imaged nine Class II disks in Taurus in the 1.3\,mm continuum
and the $J=2-1$ lines of CO, \13CO, and \C18O\ with the SMA.
CO is detected in all disks, \13CO\ in seven, and \C18O\ only in one.
To interpret the data, we created an azimuthally symmetric model
of a hydrostatically supported gas disk with a basic parameterization
of the CO chemistry incorporating dissociation at a fixed vertical
column density, $N({\rm H}_2) < 1.3\times 10^{21}$\,cm\ee,
and freeze-out below a fixed temperature, 20\,K.
Each model is characterized by nine parameters,
incorporating the stellar and disk mass, density and
temperature structure, viewing geometry, and relative \C18O\ abundance.

We found that, for disks with more than a Jupiter mass of gas,
the amount of dissociation and freeze-out is relatively minor
and most of the gas resides in a warm molecular layer that emits CO.
Model images were calculated using the radiative transfer
code, RADMC-3D.  We show that integrated line intensities correlate
with gas mass but the dispersion for any single line is very large.
The combination of \13CO\ and \C18O\ lines, however, provide a simple
and robust diagnostic of gas mass.

We compared the model grid with the observations of 6 well studied,
bright disks and found good agreement with independent mass estimates
from detailed modeling for most of them.
When applied to our SMA survey, we find these more typical disks have
low gas masses, at most a few Jupiters and many below a Jupiter.
The implied gas-to-dust ratios have a mean value of 16, with a wide
range but all well below the ISM value of 100, signifying
strong evolution in the disk composition.
The \C18O\ emission is weaker than expected for ISM abundances,
and the fits indicate a preference for a higher [CO]/[\C18O] ratio
which may be explained by selective photodissociation.

There is a remaining ambiguity in the CO abundance relative to H$_2$
in the warm molecular layer of the disks.  If this is not greatly
different than in molecular clouds and cores, the low disk masses
and low gas-to-dust ratios that we have inferred here
indicate preferential loss of gas relative to dust.
Giant planet formation should be rare in Class II disks around low mass stars.
However, if in fact the gas-to-dust ratio in disks is $\sim 100$
then the CO abundance would need to be significantly lower than in
clouds and cores, implying a unique chemistry that will be very
interesting to explore further.

Measuring the gas mass is of fundamental importance for understanding
circumstellar disk evolution and planet formation.  We have presented
a parametric modeling approach that provides a simple look-up table
for interpreting observations of low lying transitions of CO isotopologues.
We expect such observations to become much more common in the ALMA era,
and demographics of the gas content in disks to be a growing field of study
with many new results of broad impact.

\acknowledgments
We acknowledge an erudite referee report that significantly improved
the paper. We also thank Ewine van Dishoeck, Karin Oberg,
Sean Andrews, Ted Bergin, and Inga Kamp for many helpful comments and
advice that guided us from the inception to the final writeup of this idea.
This work is supported by funding from the NSF through grant AST-1208911.
We made use of the SIMBAD database, operated at CDS, Strasbourg, France,
the Exoplanet Orbit Database and the Exoplanet Data Explorer at exoplanets.org
and Astropy, a community-developed core Python package for Astronomy
\citep{2013A&A...558A..33A}.

{\it Facilities:} \facility{SMA}.

\clearpage
% REFERENCES
%\bibliographystyle{apj}
%\bibliography{references,apj-jour}

\begin{thebibliography}{81}
\expandafter\ifx\csname natexlab\endcsname\relax\def\natexlab#1{#1}\fi

\bibitem[{{Aikawa} {et~al.}(1997){Aikawa}, {Umebayashi}, {Nakano}, \&
  {Miyama}}]{1997ApJ...486L..51A}
{Aikawa}, Y., {Umebayashi}, T., {Nakano}, T., \& {Miyama}, S.~M. 1997, \apjl,
  486, L51

\bibitem[{{Aikawa} {et~al.}(2002){Aikawa}, {van Zadelhoff}, {van Dishoeck}, \&
  {Herbst}}]{2002A&A...386..622A}
{Aikawa}, Y., {van Zadelhoff}, G.~J., {van Dishoeck}, E.~F., \& {Herbst}, E.
  2002, \aap, 386, 622

\bibitem[{{Alexander} {et~al.}(2013){Alexander}, {Pascucci}, {Andrews},
  {Armitage}, \& {Cieza}}]{2013arXiv1311.1819A}
{Alexander}, R., {Pascucci}, I., {Andrews}, S., {Armitage}, P., \& {Cieza}, L.
  2013, ArXiv e-prints

\bibitem[{{All{\`e}gre} {et~al.}(2001){All{\`e}gre}, {Manh{\`e}s}, \&
  {Lewin}}]{2001E&PSL.185...49A}
{All{\`e}gre}, C., {Manh{\`e}s}, G., \& {Lewin}, {\'E}. 2001, Earth and
  Planetary Science Letters, 185, 49

\bibitem[{{Andrews} {et~al.}(2013){Andrews}, {Rosenfeld}, {Kraus}, \&
  {Wilner}}]{2013ApJ...771..129A}
{Andrews}, S.~M., {Rosenfeld}, K.~A., {Kraus}, A.~L., \& {Wilner}, D.~J. 2013,
  \apj, 771, 129

\bibitem[{{Andrews} \& {Williams}(2005)}]{2005ApJ...631.1134A}
{Andrews}, S.~M., \& {Williams}, J.~P. 2005, \apj, 631, 1134

\bibitem[{{Andrews} \& {Williams}(2007)}]{2007ApJ...671.1800A}
---. 2007, \apj, 671, 1800

\bibitem[{{Andrews} {et~al.}(2011){Andrews}, {Wilner}, {Espaillat}, {Hughes},
  {Dullemond}, {McClure}, {Qi}, \& {Brown}}]{2011ApJ...732...42A}
{Andrews}, S.~M., {Wilner}, D.~J., {Espaillat}, C., {Hughes}, A.~M.,
  {Dullemond}, C.~P., {McClure}, M.~K., {Qi}, C., \& {Brown}, J.~M. 2011, \apj,
  732, 42

\bibitem[{{Andrews} {et~al.}(2009){Andrews}, {Wilner}, {Hughes}, {Qi}, \&
  {Dullemond}}]{2009ApJ...700.1502A}
{Andrews}, S.~M., {Wilner}, D.~J., {Hughes}, A.~M., {Qi}, C., \& {Dullemond},
  C.~P. 2009, \apj, 700, 1502

\bibitem[{{Andrews} {et~al.}(2012){Andrews}, {Wilner}, {Hughes}, {Qi},
  {Rosenfeld}, {{\"O}berg}, {Birnstiel}, {Espaillat}, {Cieza}, {Williams},
  {Lin}, \& {Ho}}]{2012ApJ...744..162A}
{Andrews}, S.~M., {et~al.} 2012, \apj, 744, 162

\bibitem[{{Astropy Collaboration} {et~al.}(2013){Astropy Collaboration},
  {Robitaille}, {Tollerud}, {Greenfield}, {Droettboom}, {Bray}, {Aldcroft},
  {Davis}, {Ginsburg}, {Price-Whelan}, {Kerzendorf}, {Conley}, {Crighton},
  {Barbary}, {Muna}, {Ferguson}, {Grollier}, {Parikh}, {Nair}, {Unther},
  {Deil}, {Woillez}, {Conseil}, {Kramer}, {Turner}, {Singer}, {Fox}, {Weaver},
  {Zabalza}, {Edwards}, {Azalee Bostroem}, {Burke}, {Casey}, {Crawford},
  {Dencheva}, {Ely}, {Jenness}, {Labrie}, {Lim}, {Pierfederici}, {Pontzen},
  {Ptak}, {Refsdal}, {Servillat}, \& {Streicher}}]{2013A&A...558A..33A}
{Astropy Collaboration} {et~al.} 2013, \aap, 558, A33

\bibitem[{{Beckwith} \& {Sargent}(1991)}]{1991ApJ...381..250B}
{Beckwith}, S.~V.~W., \& {Sargent}, A.~I. 1991, \apj, 381, 250

\bibitem[{{Beckwith} \& {Sargent}(1993)}]{1993ApJ...402..280B}
---. 1993, \apj, 402, 280

\bibitem[{{Bergin} {et~al.}(2013){Bergin}, {Cleeves}, {Gorti}, {Zhang},
  {Blake}, {Green}, {Andrews}, {Evans}, {Henning}, {{\"O}berg}, {Pontoppidan},
  {Qi}, {Salyk}, \& {van Dishoeck}}]{2013Natur.493..644B}
{Bergin}, E.~A., {et~al.} 2013, \nat, 493, 644

\bibitem[{{Biller} {et~al.}(2013){Biller}, {Liu}, {Wahhaj}, {Nielsen},
  {Hayward}, {Males}, {Skemer}, {Close}, {Chun}, {Ftaclas}, {Clarke}, {Thatte},
  {Shkolnik}, {Reid}, {Hartung}, {Boss}, {Lin}, {Alencar}, {de Gouveia Dal
  Pino}, {Gregorio-Hetem}, \& {Toomey}}]{2013ApJ...777..160B}
{Biller}, B.~A., {et~al.} 2013, \apj, 777, 160

\bibitem[{{Bohlin} {et~al.}(1978){Bohlin}, {Savage}, \&
  {Drake}}]{1978ApJ...224..132B}
{Bohlin}, R.~C., {Savage}, B.~D., \& {Drake}, J.~F. 1978, \apj, 224, 132

\bibitem[{{Caselli} {et~al.}(1999){Caselli}, {Walmsley}, {Tafalla}, {Dore}, \&
  {Myers}}]{1999ApJ...523L.165C}
{Caselli}, P., {Walmsley}, C.~M., {Tafalla}, M., {Dore}, L., \& {Myers}, P.~C.
  1999, \apjl, 523, L165

\bibitem[{{Chapillon} {et~al.}(2010){Chapillon}, {Parise}, {Guilloteau},
  {Dutrey}, \& {Wakelam}}]{2010A&A...520A..61C}
{Chapillon}, E., {Parise}, B., {Guilloteau}, S., {Dutrey}, A., \& {Wakelam}, V.
  2010, \aap, 520, A61

\bibitem[{{D'Alessio} {et~al.}(2006){D'Alessio}, {Calvet}, {Hartmann},
  {Franco-Hern{\'a}ndez}, \& {Serv{\'{\i}}n}}]{2006ApJ...638..314D}
{D'Alessio}, P., {Calvet}, N., {Hartmann}, L., {Franco-Hern{\'a}ndez}, R., \&
  {Serv{\'{\i}}n}, H. 2006, \apj, 638, 314

\bibitem[{{Dame} {et~al.}(2001){Dame}, {Hartmann}, \&
  {Thaddeus}}]{2001ApJ...547..792D}
{Dame}, T.~M., {Hartmann}, D., \& {Thaddeus}, P. 2001, \apj, 547, 792

\bibitem[{{Dartois} {et~al.}(2003){Dartois}, {Dutrey}, \&
  {Guilloteau}}]{2003A&A...399..773D}
{Dartois}, E., {Dutrey}, A., \& {Guilloteau}, S. 2003, \aap, 399, 773

\bibitem[{{Dent} {et~al.}(2005){Dent}, {Greaves}, \&
  {Coulson}}]{2005MNRAS.359..663D}
{Dent}, W.~R.~F., {Greaves}, J.~S., \& {Coulson}, I.~M. 2005, \mnras, 359, 663

\bibitem[{{Dullemond} \& {Dominik}(2005)}]{2005A&A...434..971D}
{Dullemond}, C.~P., \& {Dominik}, C. 2005, \aap, 434, 971

\bibitem[{{Dullemond} {et~al.}(2002){Dullemond}, {van Zadelhoff}, \&
  {Natta}}]{2002A&A...389..464D}
{Dullemond}, C.~P., {van Zadelhoff}, G.~J., \& {Natta}, A. 2002, \aap, 389, 464

\bibitem[{{Dutrey} {et~al.}(1996){Dutrey}, {Guilloteau}, {Duvert}, {Prato},
  {Simon}, {Schuster}, \& {Menard}}]{1996A&A...309..493D}
{Dutrey}, A., {Guilloteau}, S., {Duvert}, G., {Prato}, L., {Simon}, M.,
  {Schuster}, K., \& {Menard}, F. 1996, \aap, 309, 493

\bibitem[{{Dutrey} {et~al.}(1997){Dutrey}, {Guilloteau}, \&
  {Guelin}}]{1997A&A...317L..55D}
{Dutrey}, A., {Guilloteau}, S., \& {Guelin}, M. 1997, \aap, 317, L55

\bibitem[{{Dutrey} {et~al.}(1994){Dutrey}, {Guilloteau}, \&
  {Simon}}]{1994A&A...286..149D}
{Dutrey}, A., {Guilloteau}, S., \& {Simon}, M. 1994, \aap, 286, 149

\bibitem[{{Dutrey} {et~al.}(2003){Dutrey}, {Guilloteau}, \&
  {Simon}}]{2003A&A...402.1003D}
---. 2003, \aap, 402, 1003

\bibitem[{{Favre} {et~al.}(2013){Favre}, {Cleeves}, {Bergin}, {Qi}, \&
  {Blake}}]{2013ApJ...776L..38F}
{Favre}, C., {Cleeves}, L.~I., {Bergin}, E.~A., {Qi}, C., \& {Blake}, G.~A.
  2013, \apjl, 776, L38

\bibitem[{{Frerking} {et~al.}(1982){Frerking}, {Langer}, \&
  {Wilson}}]{1982ApJ...262..590F}
{Frerking}, M.~A., {Langer}, W.~D., \& {Wilson}, R.~W. 1982, \apj, 262, 590

\bibitem[{{Gammie}(1996)}]{1996ApJ...457..355G}
{Gammie}, C.~F. 1996, \apj, 457, 355

\bibitem[{{Goldsmith} {et~al.}(1997){Goldsmith}, {Bergin}, \&
  {Lis}}]{1997ApJ...491..615G}
{Goldsmith}, P.~F., {Bergin}, E.~A., \& {Lis}, D.~C. 1997, \apj, 491, 615

\bibitem[{{Gorti} \& {Hollenbach}(2008)}]{2008ApJ...683..287G}
{Gorti}, U., \& {Hollenbach}, D. 2008, \apj, 683, 287

\bibitem[{{G{\"u}del} {et~al.}(2007){G{\"u}del}, {Padgett}, \&
  {Dougados}}]{2007prpl.conf..329G}
{G{\"u}del}, M., {Padgett}, D.~L., \& {Dougados}, C. 2007, Protostars and
  Planets V, 329

\bibitem[{{Guilloteau} {et~al.}(2013){Guilloteau}, {Di Folco}, {Dutrey},
  {Simon}, {Grosso}, \& {Pi{\'e}tu}}]{2013A&A...549A..92G}
{Guilloteau}, S., {Di Folco}, E., {Dutrey}, A., {Simon}, M., {Grosso}, N., \&
  {Pi{\'e}tu}, V. 2013, \aap, 549, A92

\bibitem[{{Hartmann} {et~al.}(1998){Hartmann}, {Calvet}, {Gullbring}, \&
  {D'Alessio}}]{1998ApJ...495..385H}
{Hartmann}, L., {Calvet}, N., {Gullbring}, E., \& {D'Alessio}, P. 1998, \apj,
  495, 385

\bibitem[{{Howard}(2013)}]{2013Sci...340..572H}
{Howard}, A.~W. 2013, Science, 340, 572

\bibitem[{{Hughes} {et~al.}(2011){Hughes}, {Wilner}, {Andrews}, {Qi}, \&
  {Hogerheijde}}]{2011ApJ...727...85H}
{Hughes}, A.~M., {Wilner}, D.~J., {Andrews}, S.~M., {Qi}, C., \& {Hogerheijde},
  M.~R. 2011, \apj, 727, 85

\bibitem[{{Hughes} {et~al.}(2007){Hughes}, {Wilner}, {Calvet}, {D'Alessio},
  {Claussen}, \& {Hogerheijde}}]{2007ApJ...664..536H}
{Hughes}, A.~M., {Wilner}, D.~J., {Calvet}, N., {D'Alessio}, P., {Claussen},
  M.~J., \& {Hogerheijde}, M.~R. 2007, \apj, 664, 536

\bibitem[{{Hughes} {et~al.}(2008){Hughes}, {Wilner}, {Qi}, \&
  {Hogerheijde}}]{2008ApJ...678.1119H}
{Hughes}, A.~M., {Wilner}, D.~J., {Qi}, C., \& {Hogerheijde}, M.~R. 2008, \apj,
  678, 1119

\bibitem[{{Jonkheid} {et~al.}(2007){Jonkheid}, {Dullemond}, {Hogerheijde}, \&
  {van Dishoeck}}]{2007A&A...463..203J}
{Jonkheid}, B., {Dullemond}, C.~P., {Hogerheijde}, M.~R., \& {van Dishoeck},
  E.~F. 2007, \aap, 463, 203

\bibitem[{{J{\o}rgensen} {et~al.}(2002){J{\o}rgensen}, {Sch{\"o}ier}, \& {van
  Dishoeck}}]{2002A&A...389..908J}
{J{\o}rgensen}, J.~K., {Sch{\"o}ier}, F.~L., \& {van Dishoeck}, E.~F. 2002,
  \aap, 389, 908

\bibitem[{{Jura} {et~al.}(2012){Jura}, {Xu}, {Klein}, {Koester}, \&
  {Zuckerman}}]{2012ApJ...750...69J}
{Jura}, M., {Xu}, S., {Klein}, B., {Koester}, D., \& {Zuckerman}, B. 2012,
  \apj, 750, 69

\bibitem[{{Kamp} \& {Dullemond}(2004)}]{2004ApJ...615..991K}
{Kamp}, I., \& {Dullemond}, C.~P. 2004, \apj, 615, 991

\bibitem[{{Kamp} {et~al.}(2011){Kamp}, {Woitke}, {Pinte}, {Tilling}, {Thi},
  {Menard}, {Duchene}, \& {Augereau}}]{2011A&A...532A..85K}
{Kamp}, I., {Woitke}, P., {Pinte}, C., {Tilling}, I., {Thi}, W.-F., {Menard},
  F., {Duchene}, G., \& {Augereau}, J.-C. 2011, \aap, 532, A85

\bibitem[{{Kastner} {et~al.}(1997){Kastner}, {Zuckerman}, {Weintraub}, \&
  {Forveille}}]{1997Sci...277...67K}
{Kastner}, J.~H., {Zuckerman}, B., {Weintraub}, D.~A., \& {Forveille}, T. 1997,
  Science, 277, 67

\bibitem[{{Lefloch} {et~al.}(1998){Lefloch}, {Castets}, {Cernicharo}, {Langer},
  \& {Zylka}}]{1998A&A...334..269L}
{Lefloch}, B., {Castets}, A., {Cernicharo}, J., {Langer}, W.~D., \& {Zylka}, R.
  1998, \aap, 334, 269

\bibitem[{{Lesniak} \& {Desch}(2011)}]{2011ApJ...740..118L}
{Lesniak}, M.~V., \& {Desch}, S.~J. 2011, \apj, 740, 118

\bibitem[{{Lynden-Bell} \& {Pringle}(1974)}]{1974MNRAS.168..603L}
{Lynden-Bell}, D., \& {Pringle}, J.~E. 1974, \mnras, 168, 603

\bibitem[{{Mathews} {et~al.}(2013){Mathews}, {Klaassen}, {Juh{\'a}sz},
  {Harsono}, {Chapillon}, {van Dishoeck}, {Espada}, {de Gregorio-Monsalvo},
  {Hales}, {Hogerheijde}, {Mottram}, {Rawlings}, {Takahashi}, \&
  {Testi}}]{2013A&A...557A.132M}
{Mathews}, G.~S., {et~al.} 2013, \aap, 557, A132

\bibitem[{{Menu} {et~al.}(2014){Menu}, {van Boekel}, {Henning}, {Chandler},
  {Linz}, {Benisty}, {Lacour}, {Min}, {Waelkens}, {Andrews}, {Calvet},
  {Carpenter}, {Corder}, {Deller}, {Greaves}, {Harris}, {Isella}, {Kwon},
  {Lazio}, {Le Bouquin}, {M{\'e}nard}, {Mundy}, {P{\'e}rez}, {Ricci},
  {Sargent}, {Storm}, {Testi}, \& {Wilner}}]{2014A&A...564A..93M}
{Menu}, J., {et~al.} 2014, \aap, 564, A93

\bibitem[{{Nomura} {et~al.}(2007){Nomura}, {Aikawa}, {Tsujimoto}, {Nakagawa},
  \& {Millar}}]{2007ApJ...661..334N}
{Nomura}, H., {Aikawa}, Y., {Tsujimoto}, M., {Nakagawa}, Y., \& {Millar}, T.~J.
  2007, \apj, 661, 334

\bibitem[{{Pani{\'c}} {et~al.}(2009){Pani{\'c}}, {Hogerheijde}, {Wilner}, \&
  {Qi}}]{2009A&A...501..269P}
{Pani{\'c}}, O., {Hogerheijde}, M.~R., {Wilner}, D., \& {Qi}, C. 2009, \aap,
  501, 269

\bibitem[{{Pavlyuchenkov} {et~al.}(2007){Pavlyuchenkov}, {Semenov}, {Henning},
  {Guilloteau}, {Pi{\'e}tu}, {Launhardt}, \& {Dutrey}}]{2007ApJ...669.1262P}
{Pavlyuchenkov}, Y., {Semenov}, D., {Henning}, T., {Guilloteau}, S.,
  {Pi{\'e}tu}, V., {Launhardt}, R., \& {Dutrey}, A. 2007, \apj, 669, 1262

\bibitem[{{Petigura} \& {Marcy}(2011)}]{2011ApJ...735...41P}
{Petigura}, E.~A., \& {Marcy}, G.~W. 2011, \apj, 735, 41

\bibitem[{{Qi} {et~al.}(2011){Qi}, {D'Alessio}, {{\"O}berg}, {Wilner},
  {Hughes}, {Andrews}, \& {Ayala}}]{2011ApJ...740...84Q}
{Qi}, C., {D'Alessio}, P., {{\"O}berg}, K.~I., {Wilner}, D.~J., {Hughes},
  A.~M., {Andrews}, S.~M., \& {Ayala}, S. 2011, \apj, 740, 84

\bibitem[{{Qi} {et~al.}(2008){Qi}, {Wilner}, {Aikawa}, {Blake}, \&
  {Hogerheijde}}]{2008ApJ...681.1396Q}
{Qi}, C., {Wilner}, D.~J., {Aikawa}, Y., {Blake}, G.~A., \& {Hogerheijde},
  M.~R. 2008, \apj, 681, 1396

\bibitem[{{Qi} {et~al.}(2006){Qi}, {Wilner}, {Calvet}, {Bourke}, {Blake},
  {Hogerheijde}, {Ho}, \& {Bergin}}]{2006ApJ...636L.157Q}
{Qi}, C., {Wilner}, D.~J., {Calvet}, N., {Bourke}, T.~L., {Blake}, G.~A.,
  {Hogerheijde}, M.~R., {Ho}, P.~T.~P., \& {Bergin}, E. 2006, \apjl, 636, L157

\bibitem[{{Qi} {et~al.}(2013){Qi}, {{\"O}berg}, {Wilner}, {D'Alessio},
  {Bergin}, {Andrews}, {Blake}, {Hogerheijde}, \& {van
  Dishoeck}}]{2013Sci...341..630Q}
{Qi}, C., {et~al.} 2013, Science, 341, 630

\bibitem[{{Ripple} {et~al.}(2013){Ripple}, {Heyer}, {Gutermuth}, {Snell}, \&
  {Brunt}}]{2013MNRAS.431.1296R}
{Ripple}, F., {Heyer}, M.~H., {Gutermuth}, R., {Snell}, R.~L., \& {Brunt},
  C.~M. 2013, \mnras, 431, 1296

\bibitem[{{Roberge} {et~al.}(2006){Roberge}, {Feldman}, {Weinberger},
  {Deleuil}, \& {Bouret}}]{2006Natur.441..724R}
{Roberge}, A., {Feldman}, P.~D., {Weinberger}, A.~J., {Deleuil}, M., \&
  {Bouret}, J.-C. 2006, \nat, 441, 724

\bibitem[{{Robitaille} {et~al.}(2006){Robitaille}, {Whitney}, {Indebetouw},
  {Wood}, \& {Denzmore}}]{2006ApJS..167..256R}
{Robitaille}, T.~P., {Whitney}, B.~A., {Indebetouw}, R., {Wood}, K., \&
  {Denzmore}, P. 2006, \apjs, 167, 256

\bibitem[{{Rosenfeld} {et~al.}(2013{\natexlab{a}}){Rosenfeld}, {Andrews},
  {Hughes}, {Wilner}, \& {Qi}}]{2013ApJ...774...16R}
{Rosenfeld}, K.~A., {Andrews}, S.~M., {Hughes}, A.~M., {Wilner}, D.~J., \&
  {Qi}, C. 2013{\natexlab{a}}, \apj, 774, 16

\bibitem[{{Rosenfeld} {et~al.}(2013{\natexlab{b}}){Rosenfeld}, {Andrews},
  {Wilner}, {Kastner}, \& {McClure}}]{2013ApJ...775..136R}
{Rosenfeld}, K.~A., {Andrews}, S.~M., {Wilner}, D.~J., {Kastner}, J.~H., \&
  {McClure}, M.~K. 2013{\natexlab{b}}, \apj, 775, 136

\bibitem[{{Rosenfeld} {et~al.}(2012){Rosenfeld}, {Qi}, {Andrews}, {Wilner},
  {Corder}, {Dullemond}, {Lin}, {Hughes}, {D'Alessio}, \&
  {Ho}}]{2012ApJ...757..129R}
{Rosenfeld}, K.~A., {et~al.} 2012, \apj, 757, 129

\bibitem[{{Semenov} \& {Wiebe}(2011)}]{2011ApJS..196...25S}
{Semenov}, D., \& {Wiebe}, D. 2011, \apjs, 196, 25

\bibitem[{{Shimajiri} {et~al.}(2014){Shimajiri}, {Kitamura}, {Saito}, {Momose},
  {Nakamura}, {Dobashi}, {Shimoikura}, {Nishitani}, {Yamabi}, {Hara},
  {Katakura}, {Tsukagoshi}, {Tanaka}, \& {Kawabe}}]{2014A&A...564A..68S}
{Shimajiri}, Y., {et~al.} 2014, \aap, 564, A68

\bibitem[{{Sun} {et~al.}(2006){Sun}, {Kramer}, {Ossenkopf}, {Bensch},
  {Stutzki}, \& {Miller}}]{2006A&A...451..539S}
{Sun}, K., {Kramer}, C., {Ossenkopf}, V., {Bensch}, F., {Stutzki}, J., \&
  {Miller}, M. 2006, \aap, 451, 539

\bibitem[{{Tafalla} {et~al.}(2002){Tafalla}, {Myers}, {Caselli}, {Walmsley}, \&
  {Comito}}]{2002ApJ...569..815T}
{Tafalla}, M., {Myers}, P.~C., {Caselli}, P., {Walmsley}, C.~M., \& {Comito},
  C. 2002, \apj, 569, 815

\bibitem[{{Thi} {et~al.}(2004){Thi}, {van Zadelhoff}, \& {van
  Dishoeck}}]{2004A&A...425..955T}
{Thi}, W.-F., {van Zadelhoff}, G.-J., \& {van Dishoeck}, E.~F. 2004, \aap, 425,
  955

\bibitem[{{Thi} {et~al.}(2010){Thi}, {Mathews}, {M{\'e}nard}, {Woitke},
  {Meeus}, {Riviere-Marichalar}, {Pinte}, {Howard}, {Roberge}, {Sandell},
  {Pascucci}, {Riaz}, {Grady}, {Dent}, {Kamp}, {Duch{\^e}ne}, {Augereau},
  {Pantin}, {Vandenbussche}, {Tilling}, {Williams}, {Eiroa}, {Barrado},
  {Alacid}, {Andrews}, {Ardila}, {Aresu}, {Brittain}, {Ciardi}, {Danchi},
  {Fedele}, {de Gregorio-Monsalvo}, {Heras}, {Huelamo}, {Krivov}, {Lebreton},
  {Liseau}, {Martin-Zaidi}, {Mendigut{\'{\i}}a}, {Montesinos}, {Mora},
  {Morales-Calderon}, {Nomura}, {Phillips}, {Podio}, {Poelman}, {Ramsay},
  {Rice}, {Solano}, {Walker}, {White}, \& {Wright}}]{2010A&A...518L.125T}
{Thi}, W.-F., {et~al.} 2010, \aap, 518, L125

\bibitem[{{van der Marel} {et~al.}(2013){van der Marel}, {van Dishoeck},
  {Bruderer}, {Birnstiel}, {Pinilla}, {Dullemond}, {van Kempen}, {Schmalzl},
  {Brown}, {Herczeg}, {Mathews}, \& {Geers}}]{2013Sci...340.1199V}
{van der Marel}, N., {et~al.} 2013, Science, 340, 1199

\bibitem[{{van Dishoeck} \& {Black}(1988)}]{1988ApJ...334..771V}
{van Dishoeck}, E.~F., \& {Black}, J.~H. 1988, \apj, 334, 771

\bibitem[{{van Zadelhoff} {et~al.}(2001){van Zadelhoff}, {van Dishoeck}, {Thi},
  \& {Blake}}]{2001A&A...377..566V}
{van Zadelhoff}, G.-J., {van Dishoeck}, E.~F., {Thi}, W.-F., \& {Blake}, G.~A.
  2001, \aap, 377, 566

\bibitem[{{Visser} {et~al.}(2009){Visser}, {van Dishoeck}, \&
  {Black}}]{2009A&A...503..323V}
{Visser}, R., {van Dishoeck}, E.~F., \& {Black}, J.~H. 2009, \aap, 503, 323

\bibitem[{{Weidenschilling}(1977)}]{1977MNRAS.180...57W}
{Weidenschilling}, S.~J. 1977, \mnras, 180, 57

\bibitem[{{Williams} \& {Cieza}(2011)}]{2011ARA&A..49...67W}
{Williams}, J.~P., \& {Cieza}, L.~A. 2011, \araa, 49, 67

\bibitem[{{Wilson} \& {Rood}(1994)}]{1994ARA&A..32..191W}
{Wilson}, T.~L., \& {Rood}, R. 1994, \araa, 32, 191

\bibitem[{{Woitke} {et~al.}(2009){Woitke}, {Kamp}, \&
  {Thi}}]{2009A&A...501..383W}
{Woitke}, P., {Kamp}, I., \& {Thi}, W.-F. 2009, \aap, 501, 383

\bibitem[{{Woitke} {et~al.}(2010){Woitke}, {Pinte}, {Tilling}, {M{\'e}nard},
  {Kamp}, {Thi}, {Duch{\^e}ne}, \& {Augereau}}]{2010MNRAS.405L..26W}
{Woitke}, P., {Pinte}, C., {Tilling}, I., {M{\'e}nard}, F., {Kamp}, I., {Thi},
  W.-F., {Duch{\^e}ne}, G., \& {Augereau}, J.-C. 2010, \mnras, 405, L26

\bibitem[{{Zuckerman} \& {Song}(2004)}]{2004ARA&A..42..685Z}
{Zuckerman}, B., \& {Song}, I. 2004, \araa, 42, 685

\end{thebibliography}

\appendix
\section{SMA observing journal}
\label{sec:obslog}
\begin{deluxetable}{rlllcccc}
\tablecolumns{8}
\tabletypesize{\scriptsize}
\tablewidth{0pt}
\tablecaption{SMA Observing Journal \label{tab:obslog}}
\tablehead{
\colhead{Source} & \colhead{R.A.(2000)} & \colhead{Dec(2000)} &
\colhead{Observing dates} & \colhead{$\tau_{225}$} &
\colhead{Array\tablenotemark{a}} & \colhead{$N_{\rm ants}$} &
\colhead{$t_{\rm int}$\tablenotemark{b}}
}
\startdata
AA Tau          & 04:34:55.42 & 24:28:53.2 & 2011 Nov 23 & 0.20 & C & 8 & 1.2 \\
                &             &            & 2012 Sep 06 & 0.10 & E & 7 & 1.9 \\
                &             &            & 2012 Oct 28 & 0.20 & C & 7 & 1.9 \\
BP Tau          & 04:19:15.84 & 29:06:26.9 & 2012 Nov 01 & 0.15 & C & 7 & 2.5 \\
                &             &            & 2012 Dec 01 & 0.20 & C & 7 & 2.7 \\
                &             &            & 2012 Dec 04 & 0.20 & C & 6 & 2.6 \\
                &             &            & 2012 Dec 25 & 0.08 & E & 7 & 2.5 \\
                &             &            & 2012 Dec 27 & 0.22 & E & 7 & 1.2 \\
                &             &            & 2013 Jan 03 & 0.14 & E & 7 & 2.8 \\
CI Tau          & 04:33:52.00 & 22:50:30.2 & 2011 Dec 04 & 0.12 & C & 8 & 1.1 \\
                &             &            & 2012 Sep 10 & 0.12 & E & 7 & 1.9 \\
CY Tau          & 04:17:33.73 & 28:20:46.9 & 2010 Nov 01 & 0.05 & C & 7 & 2.7 \\
                &             &            & 2012 Sep 10 & 0.12 & E & 7 & 1.4 \\
DL Tau          & 04:33:39.06 & 25:20:38.2 & 2011 Nov 23 & 0.20 & C & 8 & 1.3 \\
                &             &            & 2012 Sep 10 & 0.12 & E & 7 & 1.2 \\
                &             &            & 2012 Nov 01 & 0.15 & C & 7 & 2.5 \\
DO Tau          & 04:38:28.58 & 26:10:49.4 & 2012 Sep 11 & 0.12 & E & 7 & 1.4 \\
                &             &            & 2012 Nov 04 & 0.13 & C & 7 & 2.2 \\
DQ Tau          & 04:46:53.04 & 17:00:00.5 & 2011 Dec 04 & 0.12 & C & 8 & 1.3 \\
                &             &            & 2012 Sep 06 & 0.10 & E & 7 & 1.6 \\
                &             &            & 2012 Oct 28 & 0.20 & C & 7 & 1.9 \\
                &             &            & 2012 Dec 01 & 0.20 & C & 7 & 2.1 \\
                &             &            & 2012 Dec 04 & 0.20 & C & 6 & 2.5 \\
Haro 6-13       & 04:32:15.41 & 24:28:59.8 & 2011 Nov 30 & 0.15 & C & 8 & 1.7 \\
                &             &            & 2012 Sep 11 & 0.12 & E & 7 & 1.2 \\
IQ Tau          & 04:29:51.56 & 26:06:44.9 & 2012 Nov 01 & 0.15 & C & 7 & 2.5 \\
                &             &            & 2012 Dec 01 & 0.20 & C & 7 & 2.4 \\
                &             &            & 2012 Dec 04 & 0.20 & C & 6 & 2.5 \\
                &             &            & 2012 Dec 25 & 0.25 & E & 7 & 2.1 \\
                &             &            & 2012 Dec 27 & 0.22 & E & 7 & 1.0 \\
                &             &            & 2013 Jan 03 & 0.14 & E & 7 & 2.5 \\
\enddata
\tablenotetext{a}{C = compact, E = extended.}
\tablenotetext{b}{On-source integration time in hours.}
\end{deluxetable}

\clearpage
\section{The effect of different vertical temperature profiles}
\label{sec:temperature}
Our parametric investigation of disk structure on CO isotopologue
line intensities considered a range of disk midplane and atmosphere
temperatures but fixed the form of the connecting function between
them, described in equation~\ref{eq:temperature}.
To assess the effect of different vertical temperature profiles,
we ran a small set of disk models that varied the hitherto fixed
parameters, $\delta$, and $z_q/H_p$. These affect, respectively,
the vertical gradient and location at which the temperature
transitions from the cold midplane to the warm atmosphere.
The main disk parameters were fixed to values
$M_{\rm star}=1\,M_\odot,
M_{\rm gas}=0.01\,M_\odot, R_{\rm c}=60\,{\rm AU}, \gamma=0.75,
T_{\rm mid,1}=200\,{\rm K}, T_{\rm atm,1}=1000\,{\rm K}, q=0.55$,
as in Figure~\ref{fig:disk_model}.

\begin{figure}[b]
\centering
\includegraphics[width=5.0in]{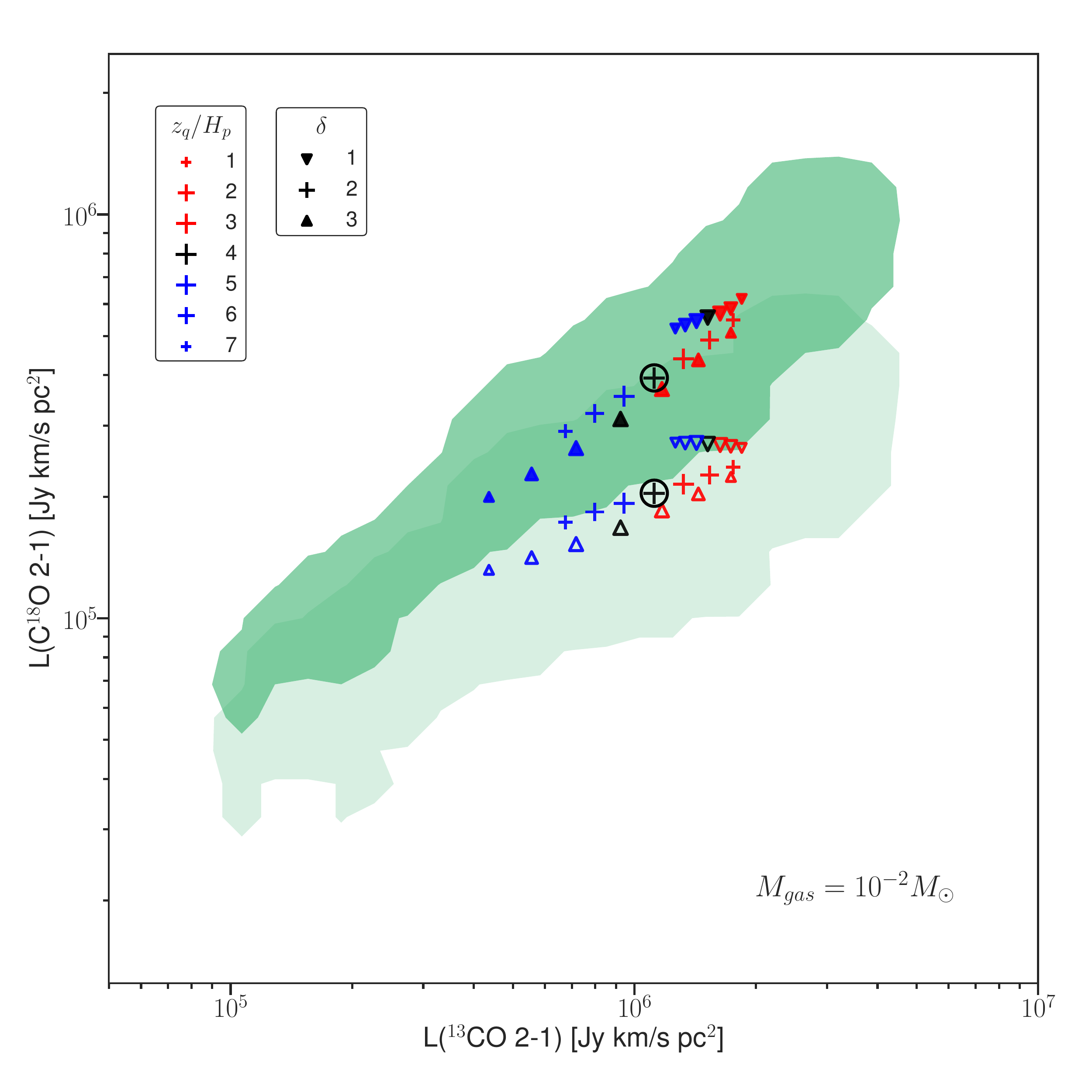}
\caption{
The effect of different vertical temperature profiles on the \13CO\
and \C18O\ 2--1 line luminosities.
The green shaded areas outline the regions within which
99\% of the full model grid points lie for disks with
$M_{\rm gas}=0.01\,M_\odot$. The lighter green area represents
the models with selective photodissociation, [CO]/[\C18O]=1650.
The different symbols, colors, and sizes indicate the vertical
temperature profile parameters, $\delta$ and $z_q/H_p$, as shown
in the legends at top left.
The two fiducial values in the full grid, $\delta=2, z_q/H_p=4$,
are circled. The variations in the line luminosities due to changes in
the vertical temperature distribution are smaller than variations due to
the range of disk parameters in Table~\ref{tab:grid}.
}
\label{fig:temperature}
\end{figure}

Figure~\ref{fig:temperature} plots the line luminosities for this
set of models on top of the full model grid results for the
same gas mass, $M_{\rm gas}=0.01\,M_\odot$.
The line luminosities change significantly
for different vertical temperature profiles, $\sim 0.6$\,dex for \13CO,
$\sim 0.2$\,dex for \C18O, but the range is much smaller than for
the full grid of disk models that incorporate a range of disk
geometries and radial temperature profiles.
The widest variation is for steep gradients, $\delta=3$,
and the transition occurring high above the midplane at $z_q/H_p=7$.
Such models have high mass fractions,
$f_{\rm freeze}$, below the CO condensation temperature and
consequently relatively low line luminosities that follow the
general correlation between \13CO\ and \C18O\ seen in the full grid.
For disk models with shallower gradients or transitions closer
to the midplane, the variation in line luminosities is much smaller.

We conclude that, unless a substantial fraction of the CO is
frozen out, variations in the vertical temperature profile do not
substantially broaden the line luminosity plot, Figure~\ref{fig:twolines}.
Over the range of likely values for protoplanetary disk properties,
the CO isotopologue line luminosities depend more strongly
on the gas mass than anything else.
Independently of uncertainties in the disk temperature structure,
therefore, the combination of \13CO\ and \C18O\ lines is a reliable
estimator of disk mass.

\end{document}